\documentclass[runningheads]{llncs}

\usepackage[table, dvipsnames]{xcolor}
\usepackage{eccv}

\usepackage{eccvabbrv}

\usepackage{graphicx}
\usepackage{booktabs}
\usepackage{comment}
\usepackage{lipsum} 
\usepackage{graphicx}
\usepackage{comment}
\usepackage{booktabs}
\usepackage{tabularx}
\usepackage{multirow}
\usepackage{xcolor,soul}
\usepackage{subcaption}
\usepackage{caption}
\usepackage{makecell}
\usepackage{adjustbox}
\usepackage{amsmath,amssymb,amsfonts}
\usepackage[inkscapelatex=false]{svg}

\usepackage{pifont}
\usepackage{xspace}

\definecolor{Gray}{gray}{0.9}
\definecolor{lgray}{gray}{0.95}
\definecolor{LightCyan}{rgb}{0.92,0.92,1}
\definecolor{lyellow}{rgb}{1,1,0.92}
\definecolor{lgreen}{rgb}{0.92,1,0.95}
\definecolor{llblue}{rgb}{0.92,0.93,0.95}
\definecolor{lred}{rgb}{1,0.85, 0.85}
\definecolor{tabhighlight}{HTML}{e5e5e5}

\usepackage{wrapfig}
\usepackage{listings}
\usepackage{colortbl}

\definecolor{cvprblue}{rgb}{0.21,0.49,0.74}
\definecolor{lblue}{rgb}{0.9,0.95,1}
\definecolor{lpurple}{rgb}{0.35,0.25,0.55}
\definecolor{lgreen}{rgb}{0.95,1,0.95}
\definecolor{sblue}{rgb}{0,0.45,1}
\definecolor{gold}{RGB}{248, 214, 99}
\definecolor{bronze}{RGB}{231, 188, 133}
\definecolor{silver}{RGB}{198, 210, 226}

\definecolor{codegreen}{rgb}{0,0.6,0}
\definecolor{codegray}{rgb}{0.5,0.5,0.5}
\definecolor{codepurple}{rgb}{0.58,0,0.82}
\definecolor{backcolour}{rgb}{0.95,0.95,0.92}
\lstdefinestyle{mystyle}{
  backgroundcolor=\color{backcolour}, commentstyle=\color{codegreen},
  keywordstyle=\color{magenta},
  numberstyle=\tiny\color{codegray},
  stringstyle=\color{codepurple},
  basicstyle=\ttfamily\footnotesize,
  breakatwhitespace=false,         
  breaklines=false,                 
  captionpos=b,                    
  keepspaces=true,                 
  numbers=left,                    
  numbersep=5pt,                  
  showspaces=false,                
  showstringspaces=false,
  showtabs=false,                  
  tabsize=2
}
\usepackage[accsupp]{axessibility}  

\usepackage[pagebackref,breaklinks,colorlinks,citecolor=eccvblue]{hyperref}

\usepackage{orcidlink}

\begin{document}

\title{AIM 2024 Challenge on Efficient Video Super-Resolution for AV1 Compressed Content}

\titlerunning{AIM 2024 Efficient Video Super-Resolution Challenge}

\author{
Marcos V. Conde\inst{1,2}$^{\dagger \ddagger}$\orcidlink{0000-0002-5823-4964} \and
Zhijun Lei\inst{3}$^\dagger$ \and 
Wen Li\inst{3}$^\dagger$ \and 
Christos Bampis\inst{4}$^\dagger$ \and \\
Ioannis Katsavounidis\inst{3}$^\dagger$ \and 
Radu Timofte\inst{1}$^\dagger$\orcidlink{0000-0002-1478-0402} \and
\\
Qing Luo \and Jie Song \and Linyan Jiang \and Haibo Lei \and Yaqing Li \and Ziqi Luo \and
Rongkang Dong \and
Cuixin Yang \and
Zongqi He \and
Jun Xiao \and
Zhe Xiao \and
Yushen Zuo \and
Zihang Lyu \and
Kin-Man Lam \and
Yuxuan Jiang \and
Jakub Nawała \and
Chen Feng \and
Fan Zhang \and
Xiaoqing Zhu \and
Joel Sole \and
David Bull \and
Jae-Hyeon Lee \and Dong-Hyeop Son \and Ui-Jin Choi \and
Mingjun Zheng \and Zhongbao Yang \and Long Sun \and Jinshan Pan \and Jiangxin Dong \and Jinhui Tang
}

\authorrunning{Conde, Lei, Li, Bampis, Katsavounidis, Timofte, et al.}

\institute{
Computer Vision Lab, CAIDAS \& IFI, University of Würzburg \and
Visual Computing Group, FTG, Sony PlayStation \and
Meta, Video Infrastructure Group \and
Netflix Inc. \\
$^\dagger$ Challenge Organizers, $^\ddagger$ Corresponding Author
\url{https://ai4streaming-workshop.github.io/}
}

\maketitle


\vspace{-2.5mm}
\begin{figure}
    \centering
    \begin{tabular}{c c c}
    \includegraphics[width=0.45\linewidth]{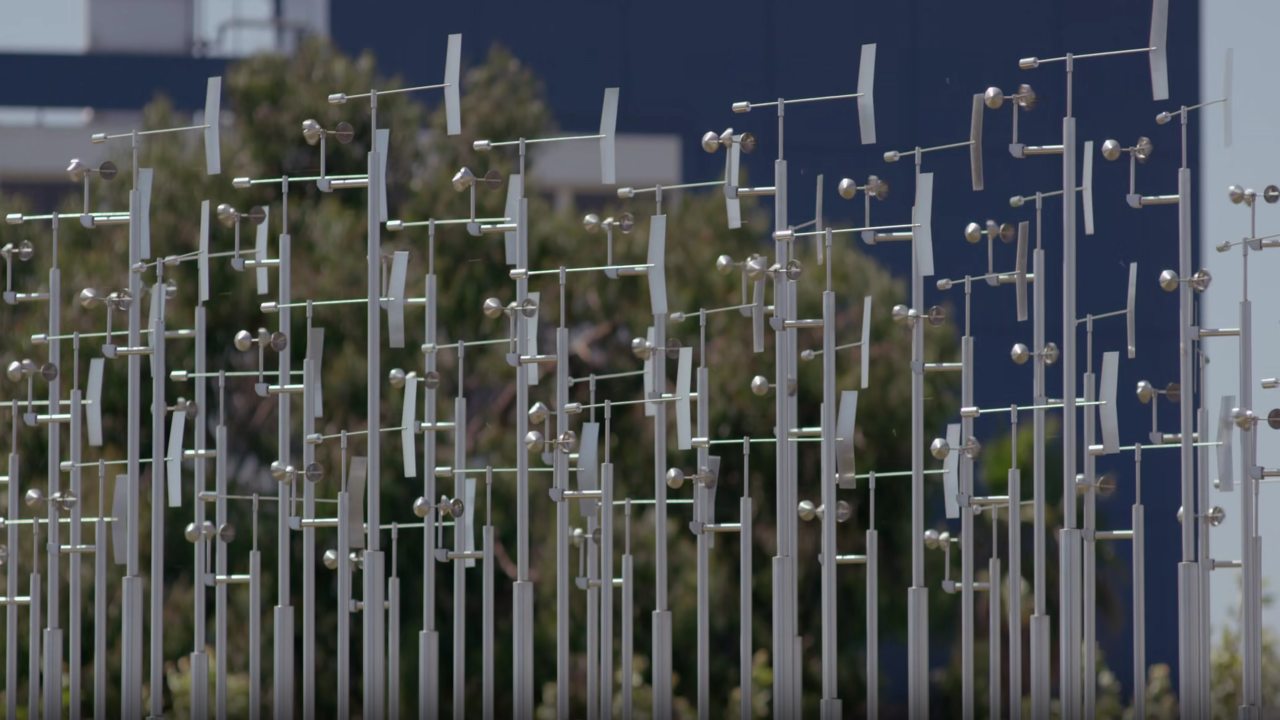} &
    \includegraphics[width=0.45\linewidth]{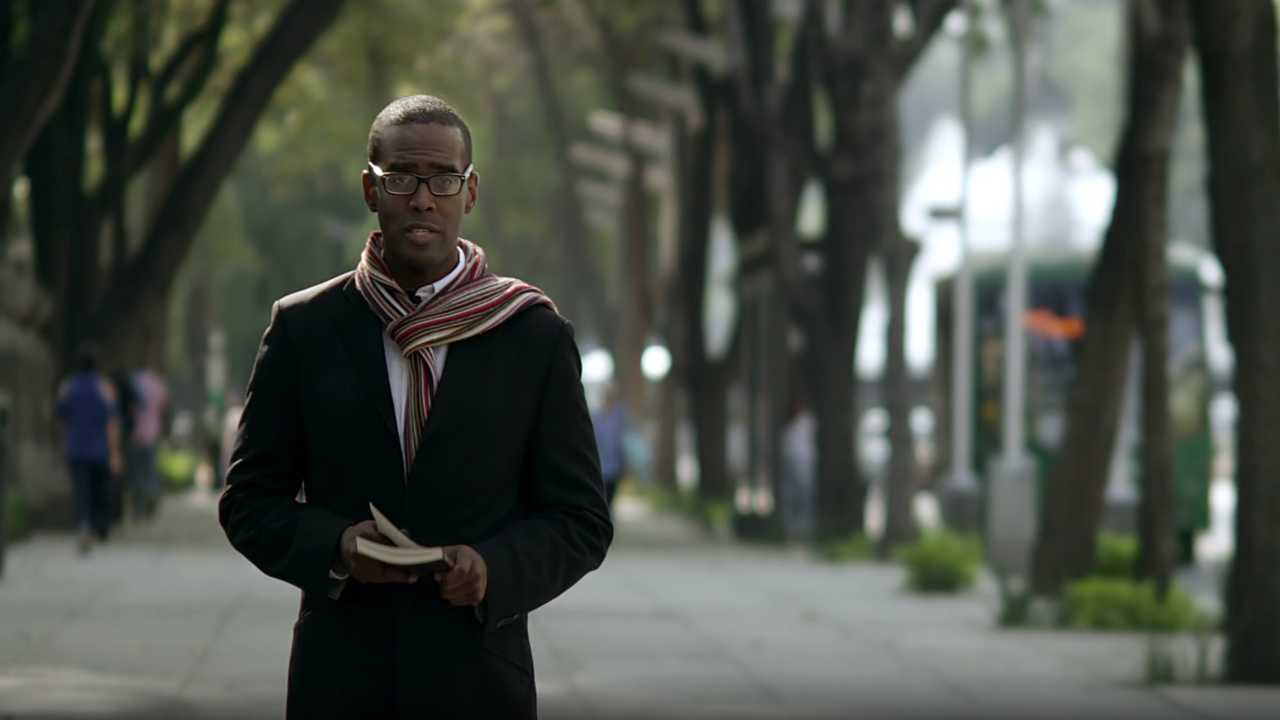} \\
    
    \includegraphics[width=0.45\linewidth]{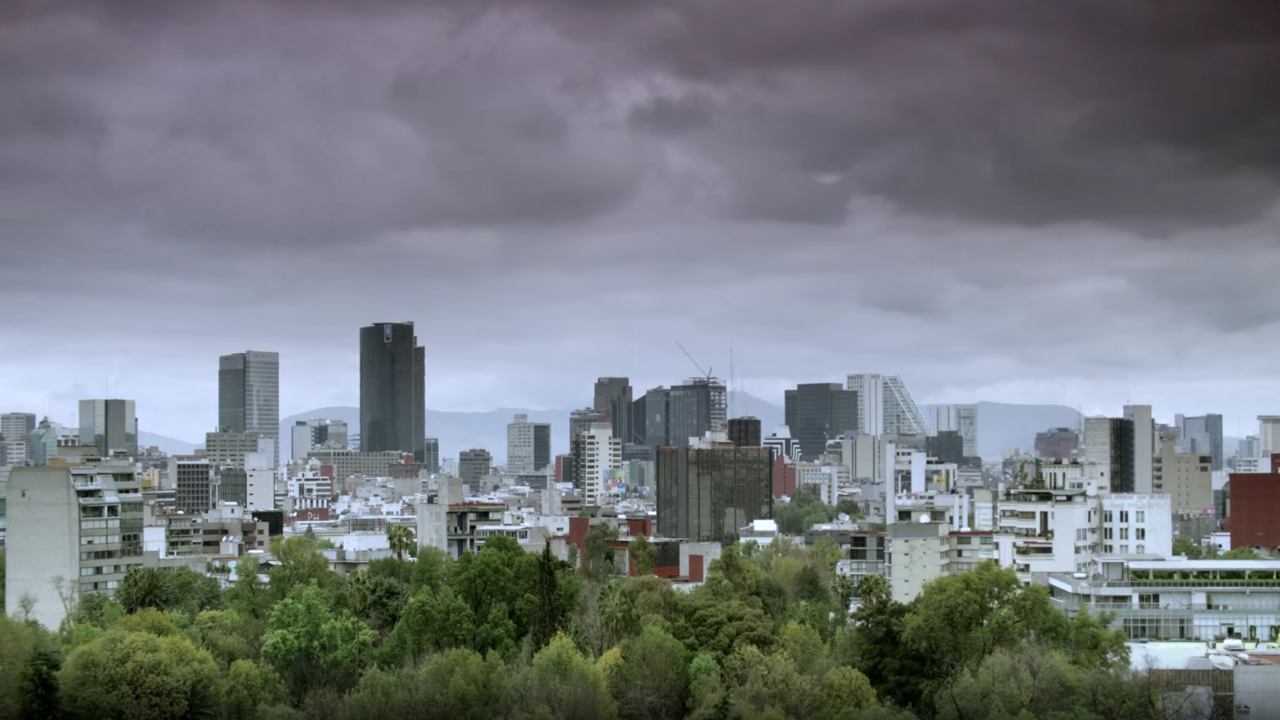} &
    \includegraphics[width=0.45\linewidth]{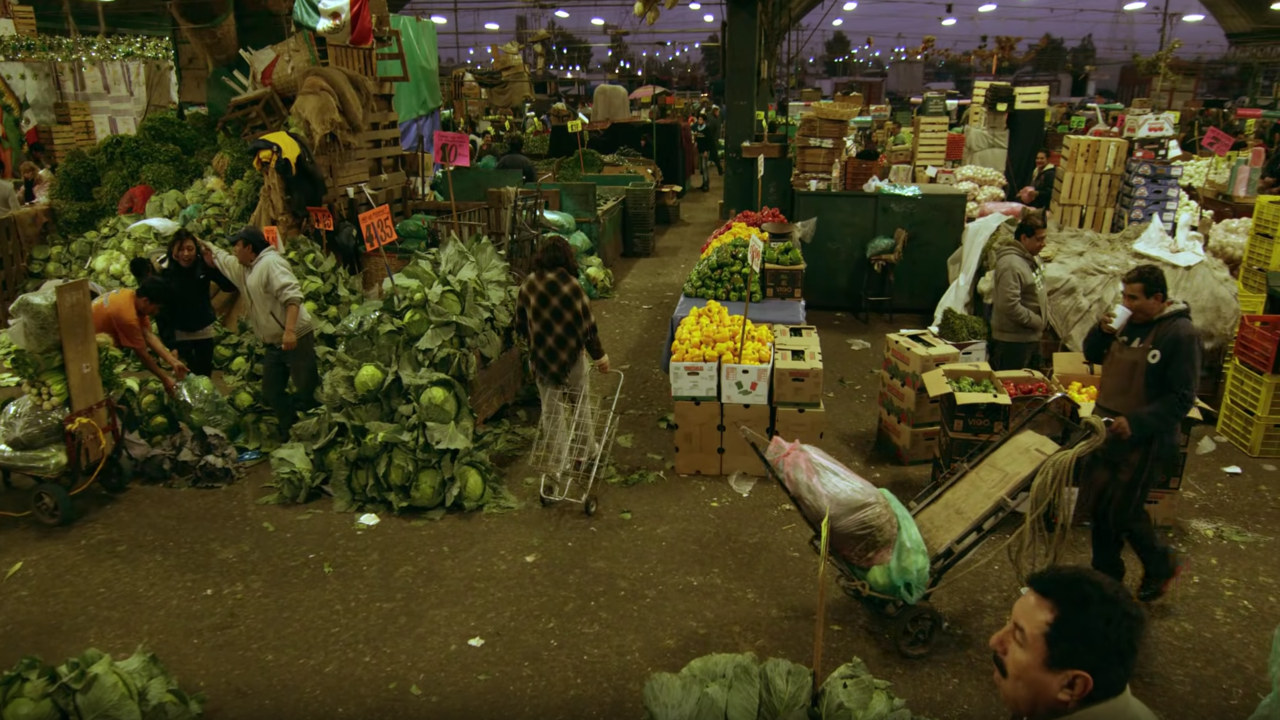} \\
    
    \includegraphics[width=0.45\linewidth]{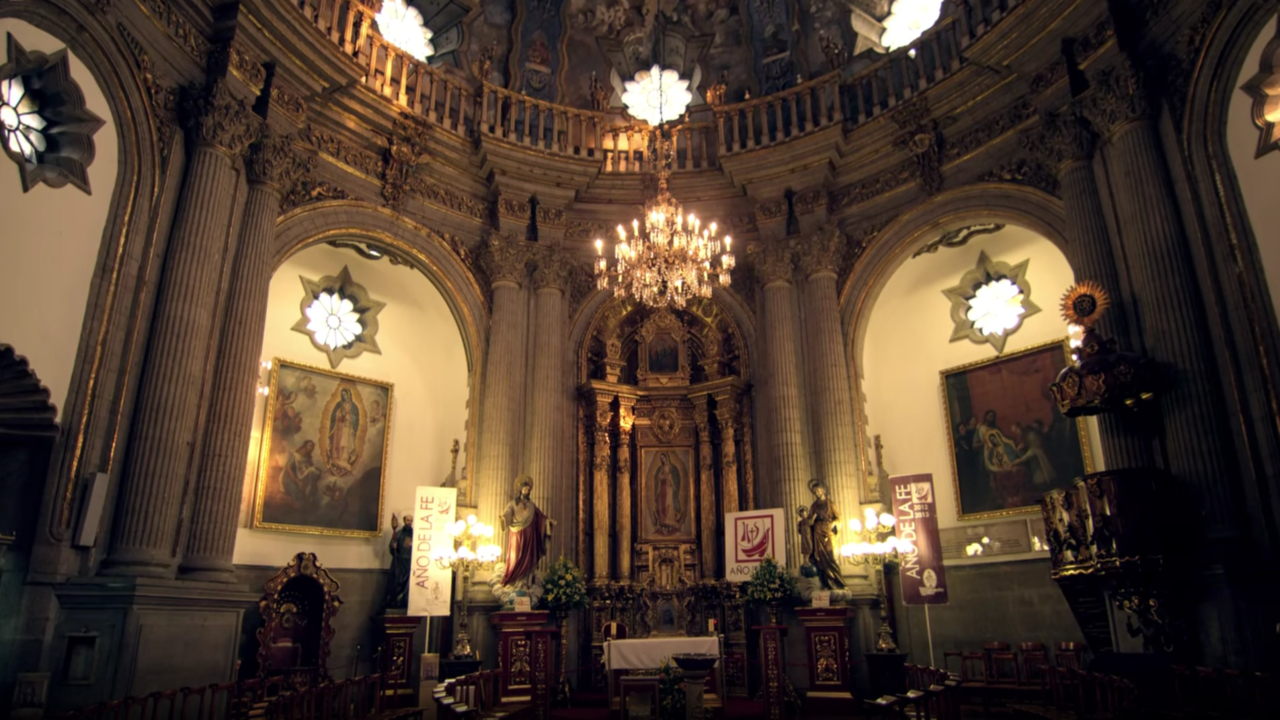} &
    \includegraphics[width=0.45\linewidth]{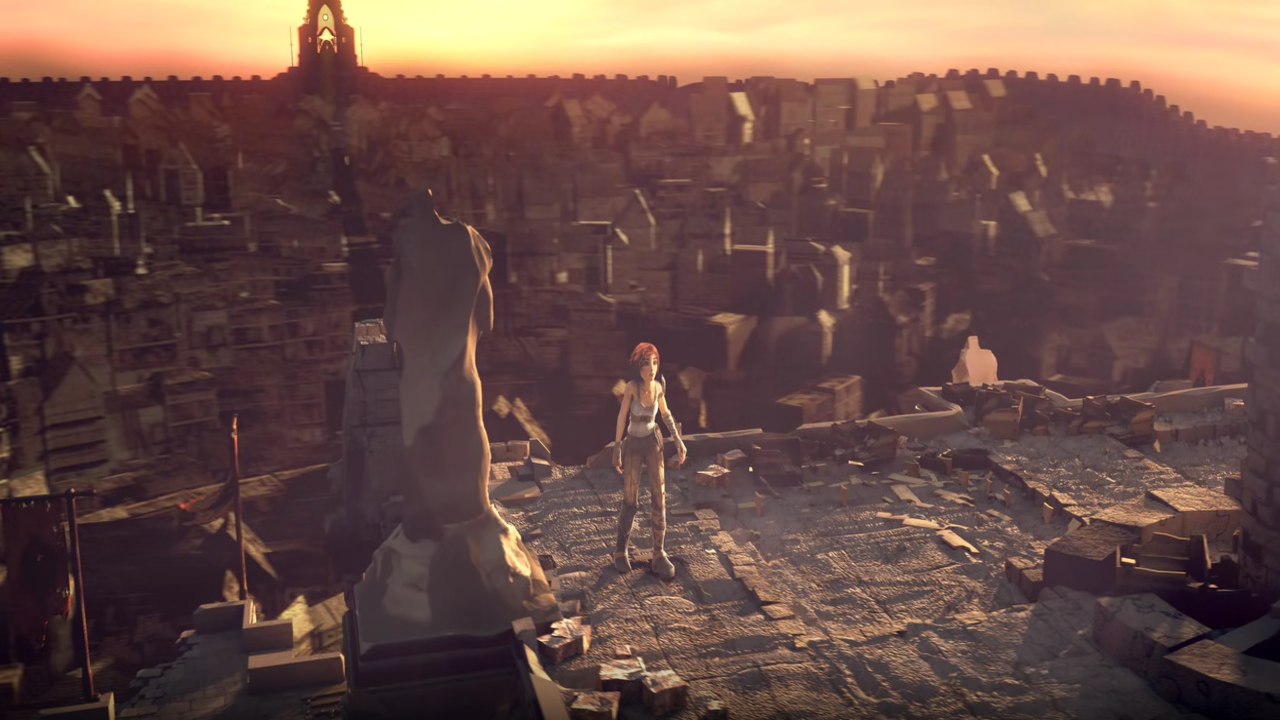} \\
    \end{tabular}
    \captionof{figure}{
    Frame samples from the high-quality test set videos \eg \emph{Netflix "El Fuente"}. The original videos are 4K resolution YCbCr 4:2:0 format.
    }
    \label{fig:teaser}
    \vspace{-7.5mm}
\end{figure}

\begin{abstract}
Video super-resolution (VSR) is a critical task for enhancing low-bitrate and low-resolution videos, particularly in streaming applications. While numerous solutions have been developed, they often suffer from high computational demands, resulting in low frame rates (FPS) and poor power efficiency, especially on mobile platforms. In this work, we compile different methods to address these challenges, the solutions are end-to-end real-time video super-resolution frameworks optimized for both high performance and low runtime. We also introduce a new test set of high-quality 4K videos to further validate the approaches. The proposed solutions tackle video up-scaling for two applications: 540p to 4K (x4) as a general case, and 360p to 1080p (x3) more tailored towards mobile devices. In both tracks, the solutions have a reduced number of parameters and operations (MACs), allow high FPS, and improve VMAF and PSNR over interpolation baselines. This report gauges some of the most efficient video super-resolution methods to date.
\end{abstract}


\section{Introduction}
\label{sec:intro}

The growing popularity of video streaming services and social media, combined with the widespread use of mobile devices, has generated a significant demand for efficient video super-resolution solutions. Over the past years, many deep learning-based solutions have been proposed for the VSR problem~\cite{ chan2021basicvsr, chan2022basicvsr++, cao2021vsr, li2023towardsvsr, shi2022rethinkingvsr}. 

The primary limitation of these methods is that they were designed to achieve high-fidelity results without being optimized for computational efficiency and mobile-related constraints, which are crucial in many real-world scenarios. For instance methods such as VideoGigaGAN~\cite{xu2024videogigagan} and BasicVSR~\cite{chan2021basicvsr} achieve outstanding results in terms of fidelity and even perceptual quality, however, these methods do not offer real-time performance (24 FPS) on regular GPUs.

Other works have proposed real-time image super-resolution solutions -- with hard memory and computational constraints -- that could be extended to the video use-cases~\cite{conde2023efficient, Conde_2024_CVPR, zamfir2023towards, yang2022aim}. We also find few challenges in the past that tackle mobile super-resolution~\cite{ignatov2021realvsr, ignatov2022powervsr}.

\vspace{2mm}
In this challenge, we take one step further in solving this problem:
\begin{enumerate}
    \item Previous works and challenges use the popular REDS~\cite{nah2019ntire} dataset, which contains $1280\times720$ 24FPS videos. Unlike previous works, we focus on 4K resolution (YCbCr 4:2:0), and we provide higher-quality videos. 
    
    \item VMAF~\cite{VMAF, li2018vmaf} was not used in previous works, although it has better correlation with the subjective perceptual quality (in comparison to simple PSNR and SSIM), for this reason, it is our main quality metric.

    \item We compress the videos using modern video codecs such as AV1~\cite{AV1}.

    \item We impose additional efficiency-related constraints on the developed solutions \ie a limit of 250 GMACs.

    \item Our challenge tackles two scenarios: 540p to 4K (x4), and 360p to 1080p (x3); each tailored for different screens and applications.
\end{enumerate}

We believe that these improvements in terms of data, evaluation and efficiency constraints, will help to push the boundaries of efficient VSR.



\paragraph{Associated AIM Challenges.} This challenge is one of the AIM 2024 Workshop\footnote{\url{https://www.cvlai.net/aim/2024/}} associated challenges on: sparse neural rendering~\cite{aim2024snr, aim2024snr_dataset}, UHD blind photo quality assessment~\cite{aim2024uhdbpqa}, compressed depth map super-resolution and restoration~\cite{aim2024cdmsrr}, efficient video super-resolution for AV1 compressed content~\cite{aim2024evsr}, video super-resolution quality assessment~\cite{aim2024vsrqa}, compressed video quality assessment~\cite{aim2024cvqa} and video saliency prediction~\cite{aim2024vsp}.

\section{Challenge}
\label{sec:challenge}

\begin{figure}[t]
    \centering
    \includegraphics[width=\linewidth]{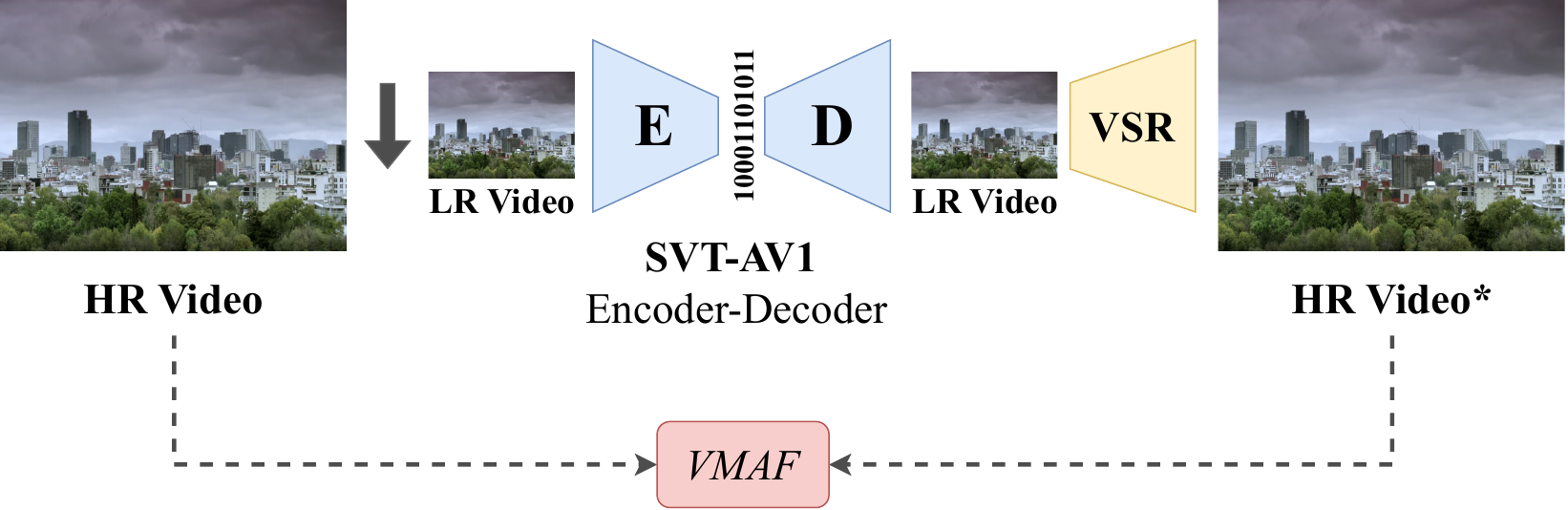}
    \caption{The challenge framework. The high-resolution (HR) videos are downscaled $\downarrow$ (with Lanczos filter) to lower resolution (LR) videos with 3x, 4x scaling ratio. We encode the videos using SVT–AV1~\cite{AV1} and different CRF values to produce encoded video bitstreams with different compression levels. We decode the videos using SVT–AV1, and we upscale the (de)compressed LR videos to the original HR resolution using the proposed video super-resolution (VSR) methods. Finally we evaluate the quality of the reconstructed HR videos (*) using well-known perceptual video quality metrics such as VMAF~\cite{VMAF, li2018vmaf}.
    }
    \label{fig:vsr_teaser}
\end{figure}

\subsection{Tracks}
\label{sec:tracks}

We consider two tracks that cover the most popular VSR applications:
\begin{enumerate}
    \item \emph{\textbf{Track 1:}} Focused on general efficient solutions, the videos are upscaled from 540p to 4K resolution (x4 scaling factor). This track extends previous work on real-time image-super resolution of compressed images~\cite{zamfir2023towards, conde2023efficient, Conde_2024_CVPR, conde2022swin2sr}.

    \item \emph{\textbf{Track 2:}} Tailored for mobile devices and small screens, the videos are upscaled from 360p to 1080p resolution (x3 scaling factor).
\end{enumerate}

\subsection{Dataset}
\label{sec:dataset}

Nineteen 4K video sequences, provided by the challenge organizer, are used to evaluate the effectiveness of the proposed super resolution solutions. These sequences contain up to 1799 frames in YCbCr 4:2:0 format. Each source video will be downscaled (with Lanczos filter) to lower resolution videos with 2x, 3x, 4x scaling ratio. 

To address real-world scenarios, we assume that the input videos have been downscaled, and also compressed. Unlike Constant Bitrate (CBR) encoding, where the bitrate is fixed throughout the video, CRF (Constant Rate Factor) allows the encoder to use as much or as little bitrate as needed to maintain a consistent quality level, thus the CRF determines the overall quality of the encoded video. CRF values typically range from 0 to 63 for the AV1 codec.

In the context of AV1 codecs~\cite{AV1}, larger Quantization Parameter (QP/CRF) values imply more compression \ie the lower the CRF value, the higher the quality of the output video. For instance, values in the range 0-20 indicate high-qualiy and low compression. In the range 50-63 the encoder compresses the video more aggressively, leading to lower quality and bitrate).

To encode the videos we use the SVT–AV1~\cite{AV1} encoder and different CRF values in the range $\{31, 39, 47, 55, 63\}$ to produce encoded video bitstreams with different compression levels. 

The encoded bitstreams will be decoded, upscaled back (using Lanczos filter as baseline) to the original resolution.

Finally, we calculate quality metrics PSNR, SSIM, and VMAF~\cite{VMAF, li2018vmaf} considering the decoded upscaled video and the ground-truth. These quality metrics will be used as reference for evaluating the super resolution proposals. Given it has been proved that VMAF has better correlation with the subjective quality, we included it as the main quality metrics for evaluating video super resolution solutions for the first time.

The library \texttt{ffmpeg}~\footnote{https://www.ffmpeg.org/download.html (version is 6.1)} was used to produce the low-resolution (LR) low-bitrate compressed videos. Bellow we provide example code:

\lstset{
    language=bash,                       
    basicstyle=\small\ttfamily,          
    keywordstyle=\color{black},          
    commentstyle=\color{black},          
    stringstyle=\color{black},           
    showstringspaces=false,              
    breaklines=true,                     
    breakatwhitespace=false,             
    postbreak=\mbox{\textcolor{black}{$\hookrightarrow$}\space}, 
    tabsize=2,                           
    frame=single,                        
    numbers=none                         
}

\begin{lstlisting}[language=bash]
ffmpeg -hide_banner -y -loglevel error -i <input> -vf 'scale=480:268:flags=lanczos+accurate_rnd+full_chroma_int:sws_dither=none:param0=5' -c:v libsvtav1 -svtav1-params preset=10:lookahead=0:keyint=-1:pred-struct=1 -crf <crf> <output> > <enc_log> 2>&1

ffmpeg -hide_banner -y -loglevel error -i <output> -i <input>  -filter_complex '[0] scale=960:536:flags=lanczos+accurate_rnd+full_chroma_int:sws_dither=none:param0=5 [enc];  [enc][1] libvmaf=feature=name=psnr|name=float_ssim:log_path=<quality_log>:log_fmt=csv' -f null - 
\end{lstlisting}

\paragraph{\textbf{Prepare pristine 1080p source videos}}

For the mobile track (360p to 1080p), the same 19 pristine 4K videos are used to produce the 1080p by cropping the source videos. Then the same Lanczos filter downscaling and encoding processes are executed to produce the compressed video bistreams with 2x, 3x scaling ratios and quality levels. 

\begin{lstlisting}[language=bash]
ffmpeg -i <input> -c:v libx264 -preset veryfast -crf 12 
-strict -2 <output.mp4>
\end{lstlisting}

\paragraph{\textbf{Evaluation}}

To evaluate the quality of the results from the proposed video super-resolution solutions, participants are requested to submit the up-scaled video bitstreams and reproducible code and models. The challenge organizers verify the same quality metrics for the 4K and 1080p source videos. The upscaled videos should be encoded with x264 encoder with near lossless encoding configuration. The following is an example code:

\begin{lstlisting}[language=Python]
# Save Upscaled Videos
imageio.mimwrite(
        "myvideo.mp4",
        output_frames,
        format="FFMPEG", codec="libx264",
        fps=input_video[2]['video_fps'],
        output_params=['-preset', 'veryfast', '-crf', 
        '12', '-strict', '2'],
        macro_block_size=None,
    )
\end{lstlisting}

\begin{lstlisting}
# Calculate Quality Metrics
ffmpeg -hide_banner -y -loglevel error -i <upscaled_video> -i <original_video> -filter_complex 'libvmaf=feature=name=psnr|name=float_ssim:log_path=<quality_result>:log_fmt=xml' -f null -
\end{lstlisting}


\section{Proposed Methods}
\label{sec:methods}

\begin{table}[t]
    \centering
    \begin{tabular}{r c c c c c}
    \toprule
    \textbf{Method} & \textbf{PSNR-Y}~$\uparrow$ & \textbf{SSIM-Y}~$\uparrow$ & \textbf{VMAF}~$\uparrow$ & \textbf{Params}~$\downarrow$ & \textbf{MACs (G)}~$\downarrow$ \\
    \midrule
    \rowcolor{lgray} \multicolumn{6}{c}{\emph{Track 1: Mobile, 360p to 1080p (X3)}} \\
    \midrule
    
    Lanczos	    & 33.123	& 0.9364	& 51.241 & - & - \\
    
    SuperBicubic++~(\ref{sec:superbic})  & 30.513	 & 0.9250	& \cellcolor{gold} 66.389 & \cellcolor{gold} 0.05 & \cellcolor{gold} 2.909 \\
    
    FSMD~(\ref{sec:FSMD}) & 32.808 & 0.9384 & \cellcolor{silver} 60.166 & \cellcolor{bronze} 1.624 & \cellcolor{bronze} 93.69 \\
    
    BVI-RTVSR~(\ref{sec:bvisr}) & 33.329 & 0.9371 & \cellcolor{bronze} 55.438 & \cellcolor{silver} 0.062 & \cellcolor{silver} 3.913 \\

    
    ETDSv2~(\ref{sec:megas}) & 32.205	& 0.9333	& 48.127 & 0.136 & 35.56 \\
    
    VPEG-VSR~(\ref{sec:vpeg})  & 28.836	& 0.8635	& 34.442 & 0.070 & 16.20 \\
    
    \midrule
    \rowcolor{lgray} \multicolumn{6}{c}{\emph{Track 2: General, 540p to 4K (X4)}} \\
    \midrule
    
    Lanczos	    & 34.651	& 0.9577	& 46.049 & - & - \\
    
    SuperBicubic++~(\ref{sec:superbic})        & 30.572	& 0.9416	& \cellcolor{gold} 64.112 & \cellcolor{silver} 0.398 & \cellcolor{silver} 206.696 \\
    
    FSMD~(\ref{sec:FSMD}) & 34.329	& 0.9591	& \cellcolor{silver} 55.920 & \cellcolor{bronze} 1.599 & \cellcolor{bronze} 207.50 \\

    BVI-RTVSR~(\ref{sec:bvisr}) & 34.464 & 0.9567 & \cellcolor{bronze} 49.829 & \cellcolor{gold} 0.063 & \cellcolor{gold} 9.595 \\
    
    
    ETDSv2~(\ref{sec:megas})   & 33.734	& 0.9564	& 43.339 & 0.136 & 35.56 \\
    
    VPEG-VSR~(\ref{sec:vpeg})    & 31.568	& 0.9111	& 36.704 & 0.077 & 20.0 \\
    
    SAFMN++~(\ref{sec:nanjing})   & 29.294	& 0.8774	& 29.225 & 0.040 & 10.22 \\
    
    \bottomrule    
    \end{tabular}
    \vspace{2mm}
    \caption{Efficient VSR Challenge Benchmark. We highlight the top-3 (gold, silver, bronze) methods that improve notably the baseline in terms of VMAF. We consider Lanczos the baseline up-sampling method. The top-3 models improve substantially Lanczos while having a limited number of MACs \ie under 250 GMACs per frame.}
    \vspace{-5mm}
    \label{tab:benchmark}
\end{table}

\begin{table*}[!ht]
    \centering
    \resizebox{\textwidth}{!}{
    \begin{tabular}{r c c c c c c c c}
        \toprule
        Method & Input & Runtime (ms) & Ensemble & \# Params. (M) & MACs (G) & GPU \\
        \midrule
        SuperBicubic++ x4~(\ref{sec:superbic})  & $960\times540$ & 10.77 & No &  0.398 & 206.7 & A100 \\
        SuperBicubic++ x3~(\ref{sec:superbic})  & $640\times360$ & 0.460 & No &  0.050 & 2.91 & A100 \\
        FSMD x4~(\ref{sec:FSMD}) & $960\times540$ & 32.33 & Yes & 1.599 & 207.5 &  4090 \\
        
        FSMD x3~(\ref{sec:FSMD}) & $640\times360$ & 13.14 & Yes & 1.624 & 93.69  &  4090 \\
        ETDSv2 x4~(\ref{sec:megas}) & $960\times540$  & 8.6 & No & 0.136 & 35.56 & A100 \\

        ETDSv2 x3~(\ref{sec:megas}) & $640\times360$  & 8.6 & No & 0.136 & 35.56 & A100 \\

        VPEG-VSR x4~(\ref{sec:vpeg}) & $960\times540$ & 8.56 & No  & 0.077 & 20.0 & 3090 \\

        VPEG-VSR x3~(\ref{sec:vpeg}) & $640\times360$ & 3.84 & No  & 0.070 & 8.10 & 3090 \\
        
        SAFMN++ x4~(\ref{sec:nanjing}) & $960\times540$ & 8.2  & No   & 0.040         & 10.22       & 3090 \\
        
        \bottomrule
    \end{tabular}
    }
    \vspace{2mm}
    \caption{Summary of implementation details for developing each solution. MACs and runtime are calculated per frame using a videoclip of 30 frames.
    }
    \vspace{-5mm}
    \label{tab:Training_specification}
\end{table*}

\paragraph{\textbf{General Ideas}} The solutions aim to improve well-known neural methods such as BasicVSR++~\cite{chan2022basicvsr++} with 5.2M parameters and $\approx 400$ GMACs per frame. Our naive baseline is the Lanczos filters that is frequently used in most video codecs.

\begin{enumerate}
    \item To ensure efficiency, all the proposed solutions process the videos in a forward manner \ie each frame independently. This simplifies the bidirectional processing used by many complex methods~\cite{chan2021basicvsr, chan2022basicvsr++} where the features of one frame can interact with past/future frame features.

    \item The neural networks are based on our previous works on real-time image super-resolution~\cite{zamfir2023towards, conde2023efficient}, especially the related work on AVIF compressed images upscaling~\cite{Conde_2024_CVPR}.
\end{enumerate}

In Table~\ref{tab:benchmark} we can compare the proposed solutions. The top-3 solutions improve notably VMAF over the baseline, they have a limited number of GMACs (always under 250), and can process each frame under 33ms, which could allow 24-30 FPS real-time upscaling. Most models have less than 150K parameters, which allows to cache (pre-load) them in memory even on mobile devices.

The solutions ETDSv2~(\ref{sec:megas}) and VPEG-VSR~(\ref{sec:vpeg}) achieved competitive performance in real-time 4K super-resolution of compressed AVIF images~\cite{Conde_2024_CVPR, sun2023safmn}. However, the properties of the videos and compression might differ notably from the image training datasets and the models do not generalize properly.

Finally, from the VMAF~\cite{VMAF} results, we could conclude that even training on single-frame (images), the models using forward processing can achieve decent temporal consistency.

\paragraph{\textbf{Summary of Implementation Details}}

A summary of the methods is provided in Table \ref{tab:Training_specification}, which includes details on the input resolution, computational complexity measured in MACs, and the number of parameters for each model.

\vspace{5mm}

In the following sections, we describe the top solutions to the challenge. Please note that the method descriptions were provided by the respective teams or individual participants as their contributions to this report.

\subsection{SuperBicubic++: An efficient and real-time super-resolution network}
\label{sec:superbic}

\emph{Qing Luo,
Jie Song,
Linyan Jiang,
Haibo Lei,
Yaqing Li,  
Ziqi Luo} \\
\textit{Tencent, China (TSR)}

\vspace{5mm}

Based on Bicubic++~\cite{bilecen2023bicubicslimslimmerslimmest} and combined with Reparameterization\cite{ding2021repvggmakingvggstyleconvnets} technology, we proposed SuperBicubic++ to improve the model effect without increasing the inference time.

In order to improve the subjective effect, based on the characteristics of human eye perception of image quality, we use the vif indicator as supervision loss for training, which effectively improves the subjective effect.

We used a three-stage training method combined with distillation learning to improve the learning effect of the small model;

\begin{table}
    \centering
    \setlength{\tabcolsep}{8pt}
    \begin{tabular}{r c c c c c}
    \toprule
    Method & ECB & Expand & Distillation & VIFLoss & VMAF~$\uparrow$     \\
    \midrule
     Bicubic++ & No & No & No & No & 69.91  \\
     A & Yes & No & No & No & 70.6561  \\
     B & Yes & Yes & No & No & 70.8431  \\
     C & Yes & Yes & Yes & No & 71.2324  \\
     SuperBicubic++ & Yes & Yes & Yes & Yes & 71.8119 \\
     \bottomrule
    \end{tabular}
    \vspace{2mm}
    \caption{SuperBicubic++ ablation results in the challenge testset.}
    \vspace{-5mm}
    \label{tab:superbic_results}
\end{table}

\subsubsection{Global Method Description}

For the 3x mobile real-time super-resolution task, Bicubic++ has a good performance in terms of speed, and the reparameterization technology can improve the model effect without increasing the inference time. Therefore, we decided to use reparameterization to replace the traditional conv3x3 to improve bicubic++. We conducted multiple experimental comparisons and finally chose the ECB Block replacement proposed in ECBSR to increase the feature extraction ability of the model; at the same time, in order to ensure the efficiency of inference, we only used 32 convolution channels, and found in the experiment that increasing the number of channels to 64 can significantly improve the performance of the model, but the time consumption will double. Therefore, we added conv1x1 during training to increase the number of channels to 64, and merged conv1x1 and conv3x3 into 32-channel conv3x3 in the inference stage, improving the model effect without changing the inference time consumption. 
For the 4x efficient super-resolution task, we chose a model structure similar to the 3x super-resolution task. However, in order to improve the model’s ability to extract features, we chose 64 channels and added more RepBlock modules.

\begin{figure}
    \centering
    \includegraphics[width=1\linewidth]{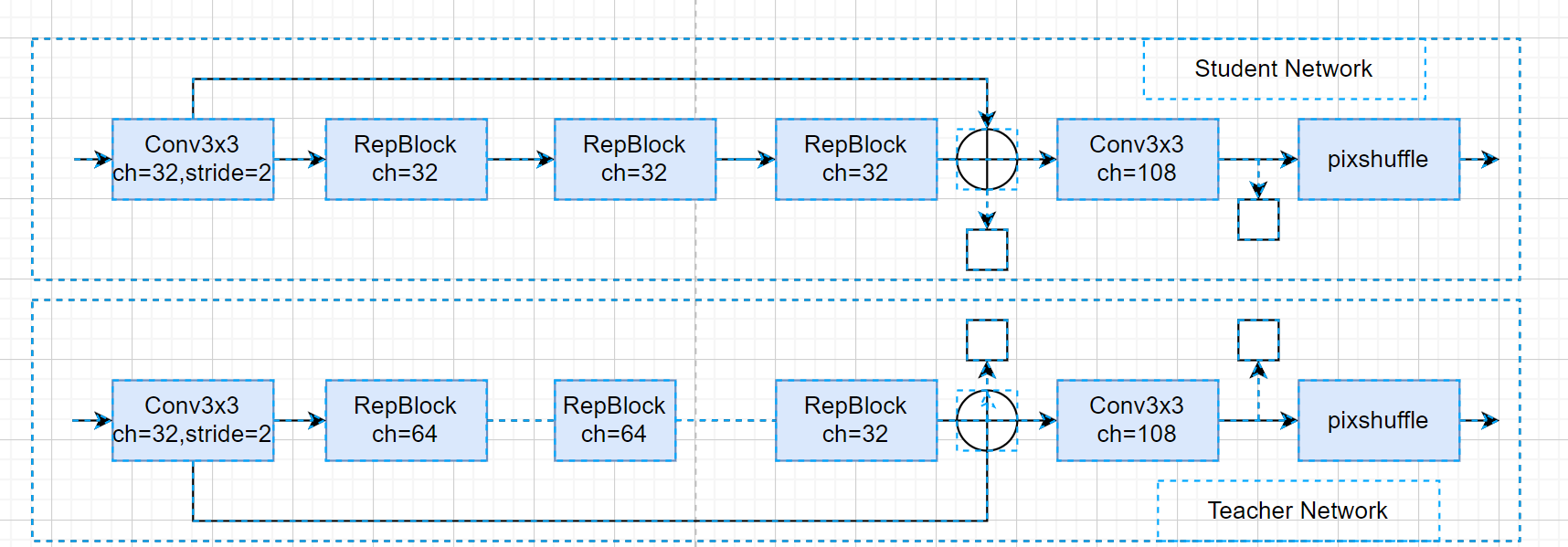}
    \caption{SuperBicubic++ X3 solution.}
    \vspace{-5mm}
    \label{fig:superbic3}
\end{figure}
\begin{figure}
    \centering
    \includegraphics[width=1\linewidth]{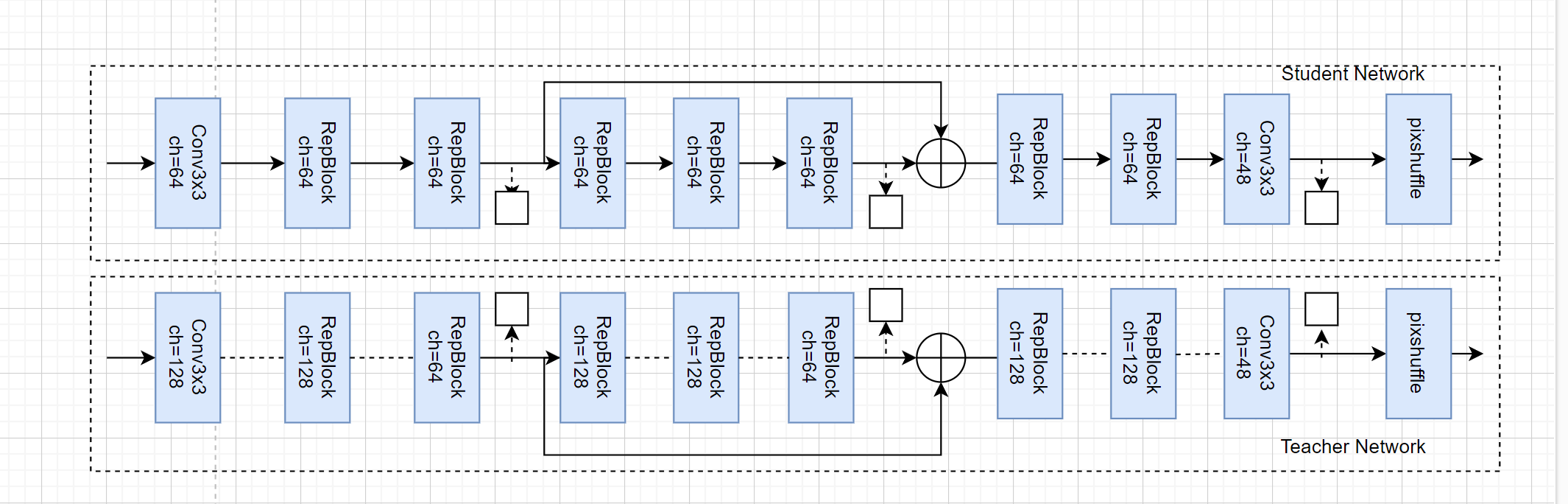}
    \caption{SuperBicubic++ X4 solution.}
    \vspace{-5mm}
    \label{fig:superbic4}
\end{figure}
\begin{figure*}[t]
    \centering
    \includegraphics[width=1\linewidth]{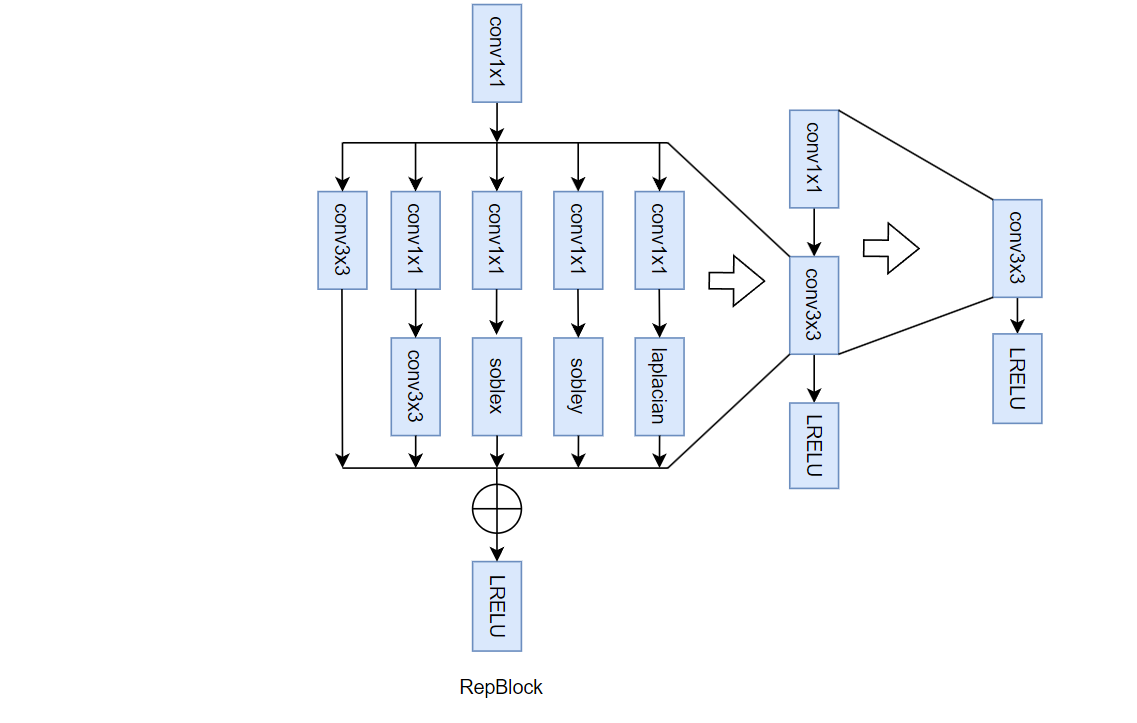}
    \label{fig:model}
    \caption{SuperBicubic++ RepBlock.}
    \vspace{-5mm}
    \label{fig:superbic_repblock}
\end{figure*}

\paragraph{\textbf{Implementation details}}

For the 3x real-time super-resolution task, In terms of data sets, we  selected the LDV3/Inter4K data set. In order to enrich the data set, we randomly used bilinear, bicubic, and lanczos sampling methods for the data set, and randomly used different crfs to encode the video, obtaining a large amount of degraded data. When reading data, we considered saving IO time and enhancing the richness of data. We flipped the input image and randomly cropped it to 128x128.
The training process is mainly divided into three stages. In the first stage, we used L1 loss, the number of iterations was 420k, and the learning rate was from 5e-4 to the minimum 5e-6; in the second stage, we used L2 loss, the number of iterations was 420k, and the learning rate was from 5e-4 to 5e-6; in the third stage, in order to improve the final subjective effect, inspired by vmaf, we found that VIF and DLM can be used as image quality evaluation indicators to reflect the quality of the image. Therefore, we conducted multiple groups of experiments and found that VIF can improve subjective image quality, but artifacts will appear when the VIF weight exceeds 0.01.
Therefore, in the third stage, L2+0.01VIF is used as loss, the number of iterations is 360K, the learning rate is from 1e-4 to the minimum 5e-6, and the cosine learning rate is used to update the learning rate in all three stages. At the same time, inspired by knowledge distillation, we use a larger model in the third stage to guide the intermediate features of our model output, helping the model to find the optimal solution and a distribution more suitable for learning.

For the 4x efficient super-resolution task, we choose a dataset processing and training process similar to 3x super-resolution.

\subsection{Fast Sequential Motion Diffusion for Real-time Video Super-resolution}
\label{sec:FSMD}

\emph{Rongkang Dong,
Cuixin Yang,
Zongqi He,
Jun Xiao,
Zhe Xiao,
Yushen Zuo,
Zihang Lyu,
Kin-Man Lam} \\
\textit{The Hong Kong Polytechnic University (POLYU-AISP)}

\vspace{5mm}

The POLYU-AISP team employs a lightweight and efficient method for video super-resolution, named Fast Sequential Motion Diffusion (FSMD). FSMD accelerates the previous recurrent neural network, TMP model \cite{zhang2023tmp}, for video super-resolution. The method incorporates Pixel-unshuffle \cite{conde2023efficient} operation to preprocess the input videos, reducing the spatial resolution of the input videos while increasing the channel dimension. This strategy enables FSMD to lower the computational load and reduce the per-frame super-resolution time, thereby achieving real-time video super-resolution.

\begin{figure*}[t]
    \centering
    \includegraphics[width=0.8\textwidth]{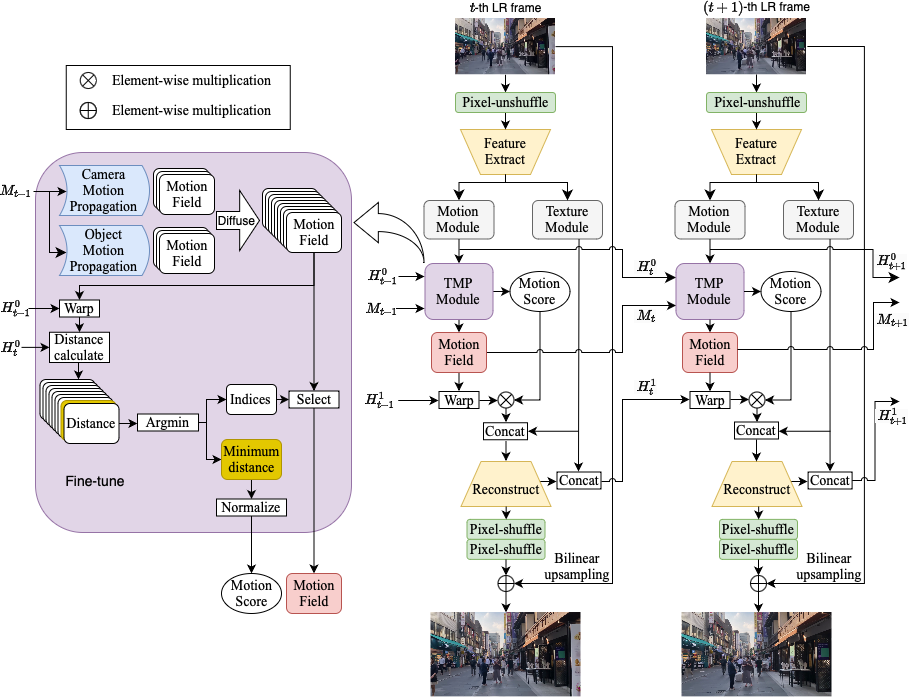}
    \caption{Team POLYU-AISP. The overall framework of the FSMD model.}
    \label{fig:polyu_framework}
\end{figure*}

The architecture of FSMD is depicted in Fig. \ref{fig:polyu_framework}. The model processes the video frames recurrently. To reduce the computational cost, we first reduce the spatial resolution of the $t$-th low-resolution (LR) frame through the pixel unshuffle operation \cite{conde2023efficient}. The $t$-th unshuffled frame is then input into the TMP model \cite{zhang2023tmp} for processing. In addition to the LR frame, the network receives the estimated motion field $M_{t-1}$ and two hidden states, $H_{t-1}^{0}$ and $H_{t-1}^{1}$, from the previous $(t-1)$-th frame. Here, $H_{t-1}^{0}$ refines the newly diffused motion field, while $H_{t-1}^{1}$ retains the texture information. The network ultimately generates a high-resolution (HR) frame, along with the updated estimated motion field $M_{t}$ and the new hidden states, $H_{t}^{0}$ and $H_{t}^{1}$.

For the Mobile Track ($\times3$), which upscales videos from 360p to 1080p, we utilize three residual blocks for feature extraction and ten for reconstruction. This configuration allows the model to restore a frame in 13.14 ms with 93.69 GFLOPs. For the Efficient Track ($\times4$), which scales videos from 540p to 4K, two residual blocks are used for feature extraction and ten for reconstruction. The model requires 32.33 ms and 207.50 GFLOPs to restore a frame.

\paragraph{Implemetation details.} We trained the FSMD model on the LDV3 dataset and the Inter4K dataset. For the LDV3 dataset, HR videos were directly downsampled with the $\times3$ and $\times4$ downscaling factors to obtain the LR videos. 
For the Inter4K dataset, we downsampled and then randomly compressed the HR videos with various compression levels, i.e., CRF=31, 39, 47, 55, 63, to obtain the LR compressed videos.
Subsequently, we extracted the HR frames and LR frames from the original HR videos and the compressed LR videos, respectively.
For each HR frame, we uniformly partitioned it into eight patches, and then applied a center crop to each patch. The resolution of the HR cropped patches is $480\times480$. The corresponding LR cropped patches were obtained using the same method, whose resolution is $160\times160$ for $\times3$ or $120\times120$ for $\times4$. This approach ensures a more balanced sampling of different regions within the frames for training.
For $\times3$, the patch size of the HR training sequence was $252\times252$. For $\times4$, the patch size of the HR training sequence was $256\times256$.
Each training sequence contained 15 frames. The batch size was set to 32, and the network was optimized using the Charbonier loss \cite{lai2017deep} and the Adam optimizer \cite{kingma2014adam} with an initial learning rate of $10^{-4}$, gradually decreasing to $10^{-6}$ using the cosine annealing scheme \cite{loshchilov2017sgdr}. The total training involved 600,000 iterations. Other experimental settings followed previous work \cite{zhang2023tmp}. Table \ref{tab:Training_specification} details the training time, the ensemble, the use of extra data, the number of parameters, MACs per frame, and the GPU used for training the FSMD model.

\textbf{Results.} A total of 19 videos were used, each downsampled at five different compression levels, resulting in 95 evaluation videos. We calculated the average VMAF, PSNR-Y, and Float-SSIM for all frames within a video and then averaged these metrics across 19 videos with different CRF values and across 95 videos. Table \ref{tab:FSMD_result} presents the results for both tracks.

\begin{table*}[h]
    \centering
    \setlength{\tabcolsep}{8pt}
    \resizebox{0.8\textwidth}{!}{
    \begin{tabular}{c c c c c}
        \toprule
        Track (CRF) & SF & VMAF & PSNR-Y & SSIM  \\
        \midrule
         Mobile (31) & $\times3$ & 85.1478 & 35.8685 & 0.9849 \\
         Mobile (39) & $\times3$ &75.9627 & 34.7595 & 0.9740 \\
         Mobile (47) & $\times3$ &62.6097 & 33.0749 & 0.9526 \\
         Mobile (55) & $\times3$ &47.8857 & 31.2279 & 0.9181 \\
         Mobile (63) & $\times3$ &29.2231 & 29.1093 & 0.8623 \\
         Mobile (all) & $\times3$ &60.1658 & 32.8080 & 0.9384 \\
         \midrule
         Efficient (31) & $\times4$ &80.6043 & 37.4050 & 0.9916 \\
         Efficient (39) & $\times4$ &70.2629 & 36.2847 & 0.9848 \\      
         Efficient (47) & $\times4$ &57.1925 & 34.6226 & 0.9709 \\
         Efficient (55) & $\times4$ &44.2841 & 32.7337 & 0.9463 \\
         Efficient (63) & $\times4$ &27.1966 & 30.5231 & 0.9019 \\
         Efficient (all) & $\times4$ &55.9081 & 34.3138 & 0.9591 \\
         \bottomrule
    \end{tabular}
    }
    \vspace{2mm}
    \caption{FSMD ablation study. Results of quality metrics for each QP value.}
    \label{tab:FSMD_result}
\end{table*}

\subsection{BVI-RTVSR: A Real-Time Video Super-Resolution Model for AV1 Compressed Content}
\label{sec:bvisr}

\emph{Yuxuan Jiang~$^1$,
Jakub Nawała~$^1$,
Chen Feng~$^1$,
Fan Zhang ~$^1$,
Xiaoqing Zhu~$^2$,
Joel Sole~$^2$,
and David Bull~$^1$} \\
\textit{$^1$ \textit{Visual Information Laboratory, University of Bristol, UK}\\
$^2$ \textit{Netflix Inc.}\\
}

\vspace{5mm}

Inspired by our previous work \cite{jiang2024mtkd, jiang2023compressing, ma2020cvegan}, we propose a low-complexity video super-resolution method to improve the visual quality of compressed video content, which specifically performs resolution up-sampling from 360p to 1080p and from 540p to 4K. The proposed approach utilizes a CNN-based network architecture, which was optimized for AV1 (SVT)-encoded content at various quantization levels. To reduce complexity, we employ Pixelunshuffle and PixelShuffle layers. Besides, we apply a Multi-teacher Knowledge-Distillation (MTKD) strategy to enhance the performance of the low-complexity model based on \cite{jiang2024mtkd, jiang2023compressing}, using the EDSR\_baseline model \cite{lim2017enhanced} and CVEGAN model \cite{ma2020cvegan} as dual-teacher. Since commonly used loss functions do not always align well with perceived quality. To this end, a perceptually inspired loss function developed in \cite{ma2020cvegan} is employed in the training and optimization processes in order to produce results with improved perceptual video quality. To increase the richness of the training data, apart from the provided LDV3 videos, original sequences from the BVI-DVC database \cite{ma2021bvi} and BVI-AOM database \cite{nawala2024bvi} are used to generate training datasets.

This approach has been tested with the SVT-AV1 version 1.8.0 video codec for evaluation and achieved an average improvement of VMAF against the provided anchor results of 4, the figure for PSNR-Y is 0.24 dB. In terms of complexity, the proposed model only performs with 3.9GMACs per frame for $\times$ 3 task and 9.6GMACs per frame for $\times$ 4 task, and an average runtime of 0.8 ms per frame for $\times$ 3 task and 2 ms for $\times$ 4 task based on RTX3090.

\begin{table}
    \setlength{\tabcolsep}{8pt}
    \centering
    \begin{tabular}{l r c c}
    \toprule
                             & Method & PSNR-Y (dB) & VMAF (score)  \\ 
    \midrule
\multirow{4}{*}{Track1 (x3)} & EDSR\_baseline & 33.73  & 57.41 \\
                             & CVEGAN & 33.69 & 57.92 \\
                             & Provided anchor & 33.14  & 51.28 \\
                             & \textbf{BVI-RTVSR}   & \textbf{33.34}  & \textbf{55.44} \\ 
    \midrule
\multirow{4}{*}{Track1 (x4)} & EDSR\_baseline & 35.32  & 52.16 \\
                             & CVEGAN & 35.30 & 53.03 \\
                             & Provided anchor & 34.66  & 45.92 \\
                             & \textbf{BVI-RTVSR}   & \textbf{34.90}  & \textbf{49.96} \\
    \bottomrule
\end{tabular}
    \vspace{2mm}
    \caption{BVI-RTVSR results summary with PSNR-Y and VMAF in average.}
    \vspace{-7mm}
    \label{tab:results_bvi}
\end{table}

\begin{table*}[]
    \centering
    \resizebox{\textwidth}{!}{
    \begin{tabular}{l cccccc}
    \toprule
                            RTX3090      & Input & Track & \makecell{Train Time\\(hrs)} & Ensemble & Extra Data \\ \hline
        \multirow{2}{*}{Training} & (48,48,3) & $\times$3 & 100 & No & \makecell{BVI-DVC\cite{ma2021bvi}\\BVI-AOM\cite{nawala2024bvi}} \\
                            & (48,48,3) & $\times$4 & 100 & No & \makecell{BVI-DVC\cite{ma2021bvi}\\BVI-AOM\cite{nawala2024bvi}} \\ 
                            
    \midrule \midrule

                                  RTX3090      & Input & Track & \makecell{\#Params.\\(M)} & \makecell{MACs\\(G/frame\textbar K/pixel)} & \makecell{RT\\(ms/frame)} \\ \hline
         \multirow{2}{*}{Testing} & (640,360,3) & $\times$3 & 0.062 & 3.913\textbar 1.887 & 0.8 \\
                                  & (960,540,3) & $\times$4 & 0.063 & 9.595\textbar 1.157 & 2 \\
    \bottomrule
    \end{tabular}
    }
    \vspace{2mm}
    \caption{BVI-RTVSR training and testing configuration and model complexity overview obtained by using the recommended tool \cite{VideoAI}.}    
    \vspace{-5mm}
    \label{tab:ablation_bvi}
\end{table*}

\subsubsection{Employed Network Architecture}

\begin{figure}[!ht]
    \begin{minipage}[t]{0.92\linewidth}
    \centering
    \centerline{\includegraphics[width=1\linewidth]{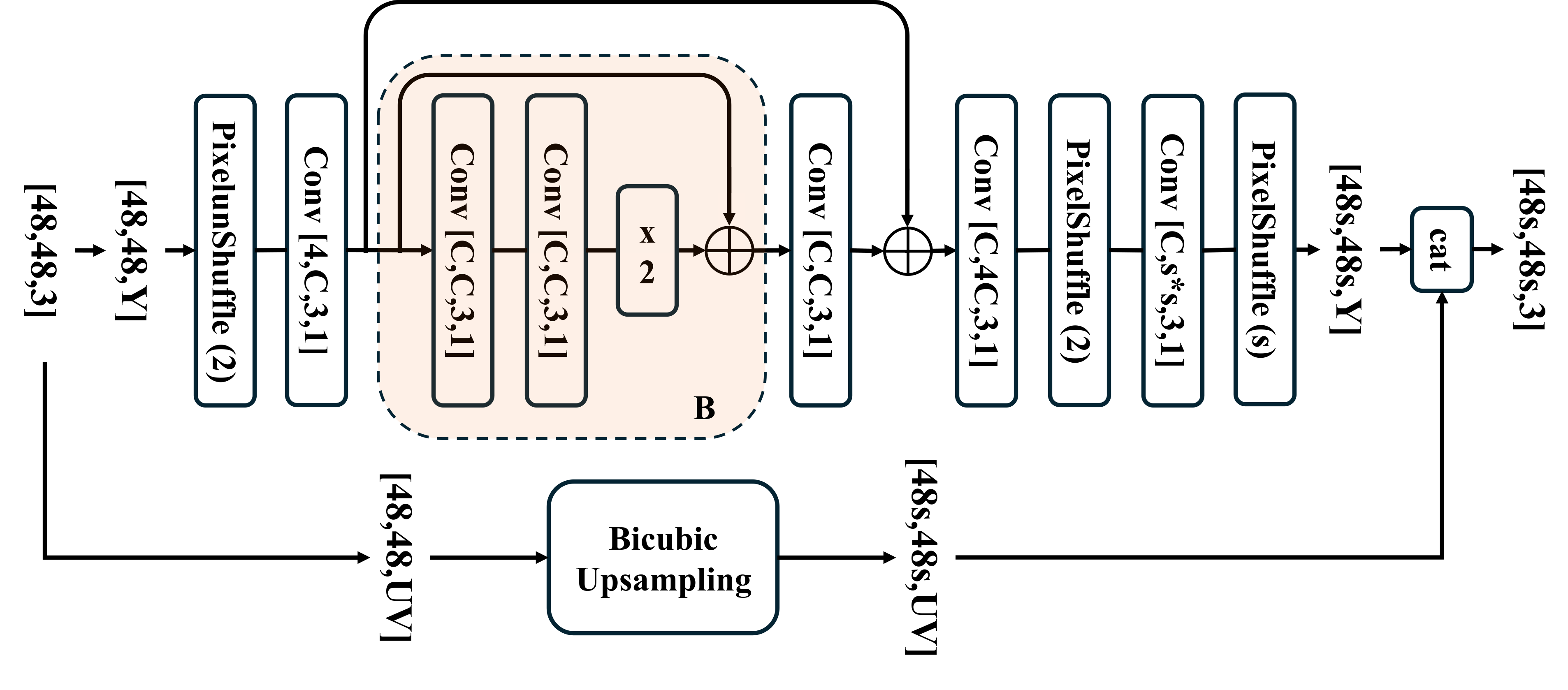}}
    \text{\small (a)}
    \end{minipage}
    
    \begin{minipage}[t]{0.92\linewidth}
    \centering
    \centerline{\includegraphics[width=1\linewidth]{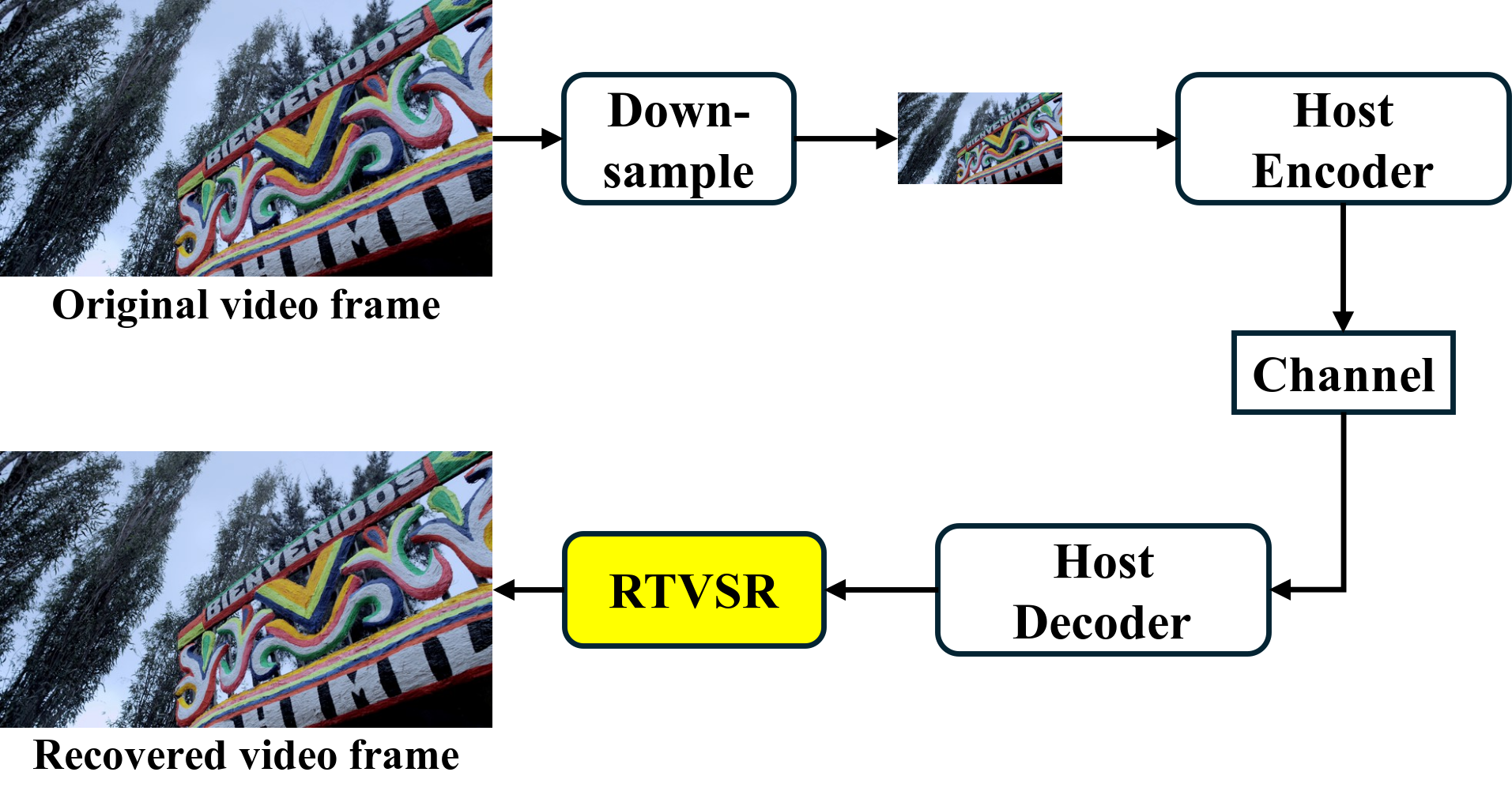}}
    \text{\small (b)}
    \end{minipage}

    \caption{\small (a) The network structure of the proposed BVI-RTVSR model. (b) The proposed coding framework, with an RTVSR module. The displayed figure targets $\times$3 upsampling, and we use the same framework and model for $\times$4 upsampling but adjusted accordingly. In (a) B equals 3, and C equals 24.}
    \label{fig:coding_frame}
\end{figure}

The network architecture (inspired by EDSR \cite{lim2017enhanced}) is shown in Figure~\ref{fig:coding_frame}(a). For the training, the 48$\times$48 YCbCr 4:2:0 compressed image block is firstly upsampled by the nearest neighbour (NN) filter to 48$\times$48 YCbCr 4:4:4 before it is fed as input to the model. The output is a processed 144$\times$144 image block (i.e., three times upsampled with respect to the input) in the format of YCbCr 4:4:4 and then converted into YCbCr 4:2:0 format, targeting its original uncompressed full resolution version. For $\times$4 task, upsampling factor 4 is used in the final PixelShuffle layer, and 192$\times$192 image block is output. Since knowledge distillation (KD) has emerged as a promising technique in deep learning \cite{he2020fakd,jiang2024mtkd}, the proposed RTVSR model is used as a student model. For the teacher model, a perceptually-inspired network for compressed video enhancement, CVEGAN \cite{ma2020cvegan} and EDSR\_baseline \cite{lim2017enhanced}, have been used.

In order to meet the real-time requirements, a PixelunShuffle layer is used before convolutional layers to significantly reduce the number of operations. The main body consists of B identical blocks, and there is a ReLU layer after two consecutive convolutional layers. The upsampling module consists of a two-step PixelShuffle operation. UV channel is upsampled by a bicubic filter and concatenated with the Y-channel recovered by the CNN model.

\subsubsection{Training Content}
\begin{figure*}[t]
    \scriptsize
    \centering
    \begin{minipage}[t]{0.3\linewidth}
        \centering
        \centerline{\includegraphics[width=.98\linewidth]{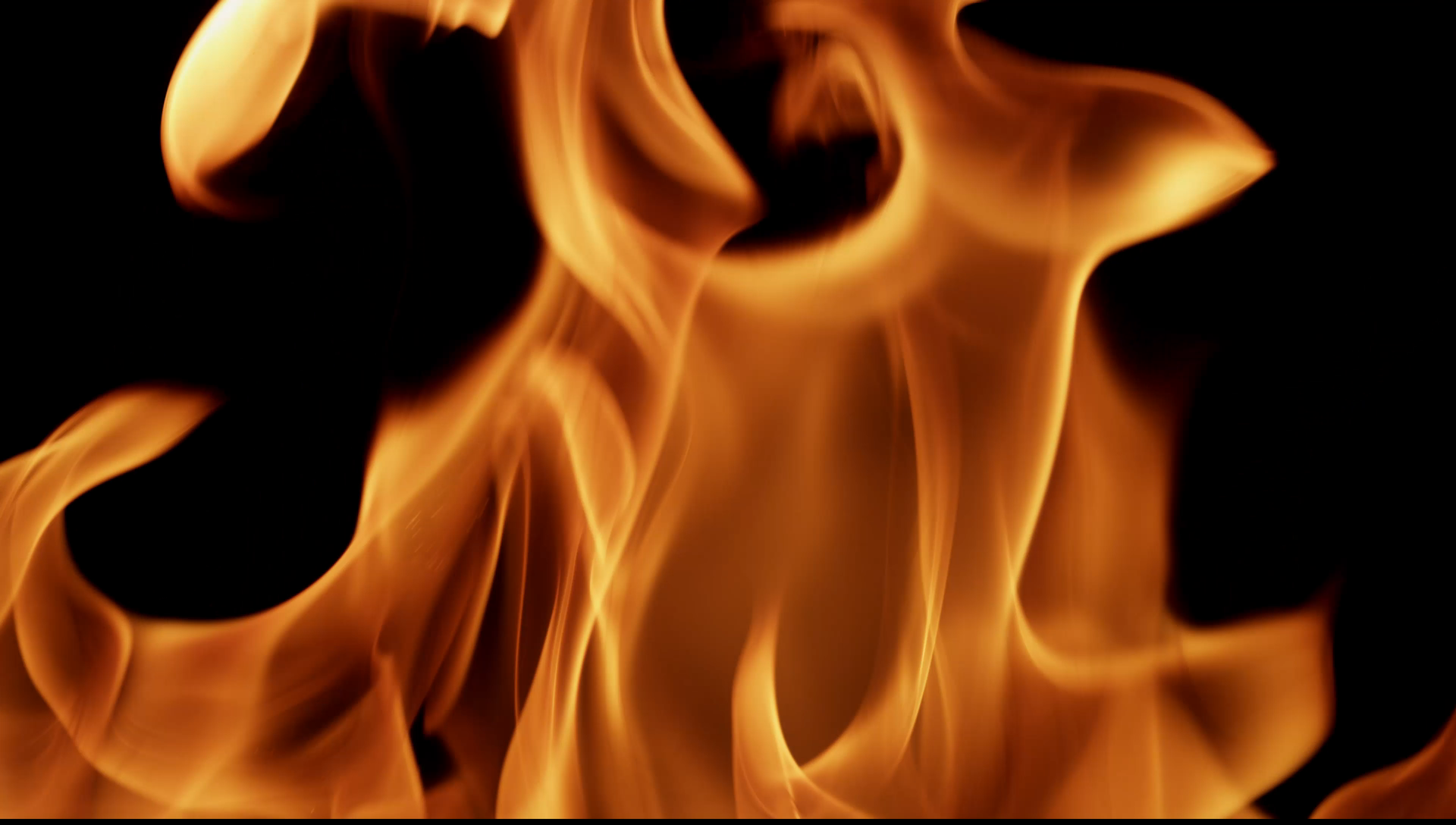}}
        \text{BVI-DVC:BFireS18Mitch}\vspace{.1cm}
    \end{minipage}
    \begin{minipage}[t]{0.3\linewidth}
        \centering
        \centerline{\includegraphics[width=.98\linewidth]{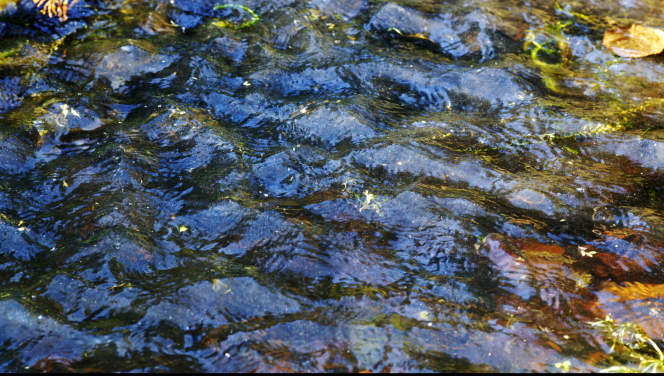}}
        \text{BVI-DVC:BCalmingWater}\vspace{.1cm}
            \end{minipage}
    \begin{minipage}[t]{0.3\linewidth}
        \centering
        \centerline{\includegraphics[width=.98\linewidth]{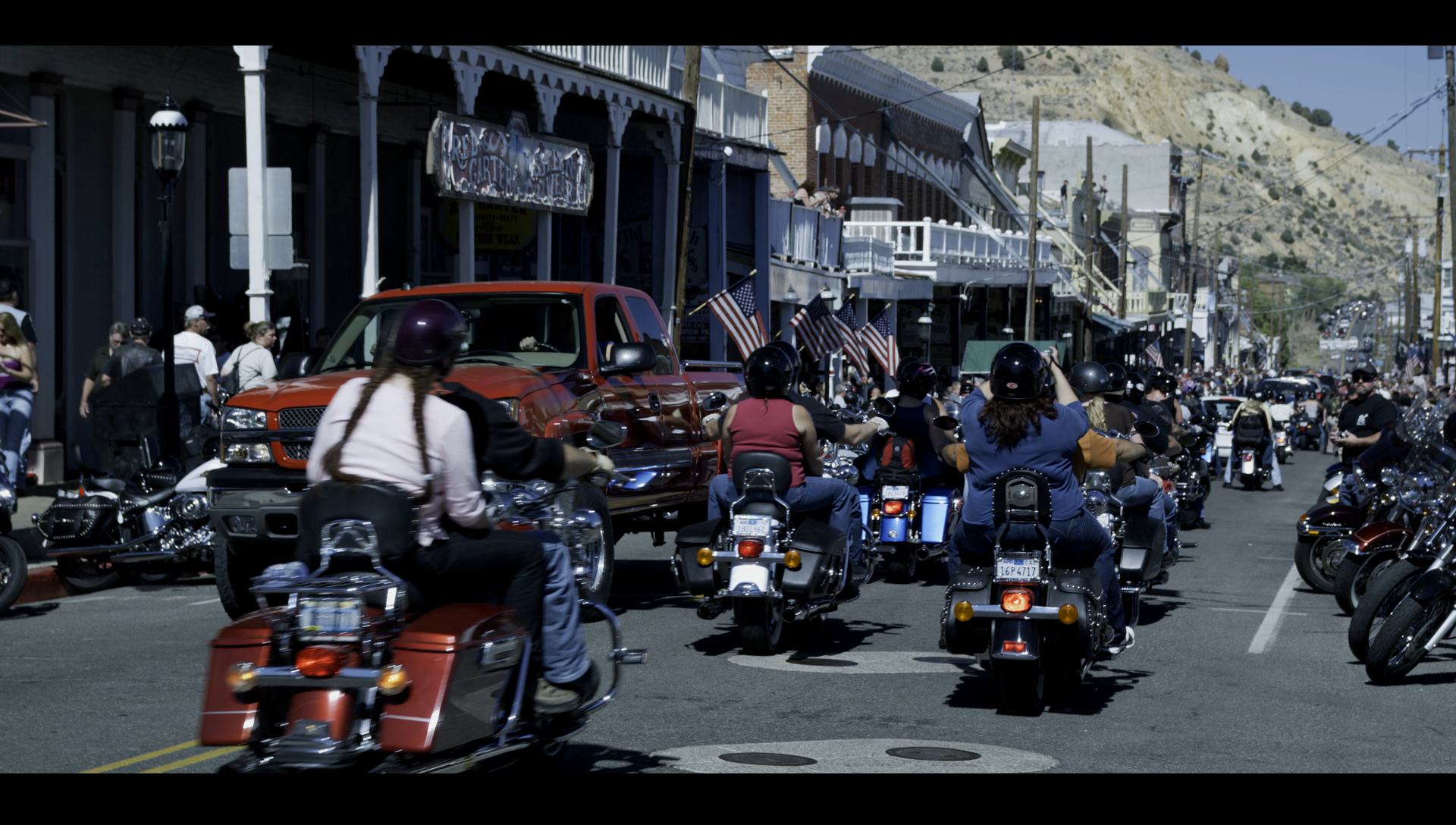}}
        \text{BVI-AOM:BHarleyDavidson}\vspace{.1cm}
        \end{minipage}

    \centering
    \begin{minipage}[t]{0.3\linewidth}
        \centering
        \centerline{\includegraphics[width=.98\linewidth]{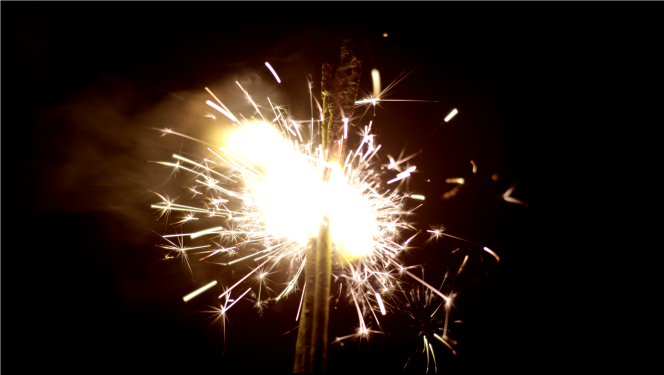}}
        \text{BVI-AOM:ASparklerBVIHFR}\vspace{.1cm}
            \end{minipage}
    \begin{minipage}[t]{0.3\linewidth}
        \centering
        \centerline{\includegraphics[width=.98\linewidth]{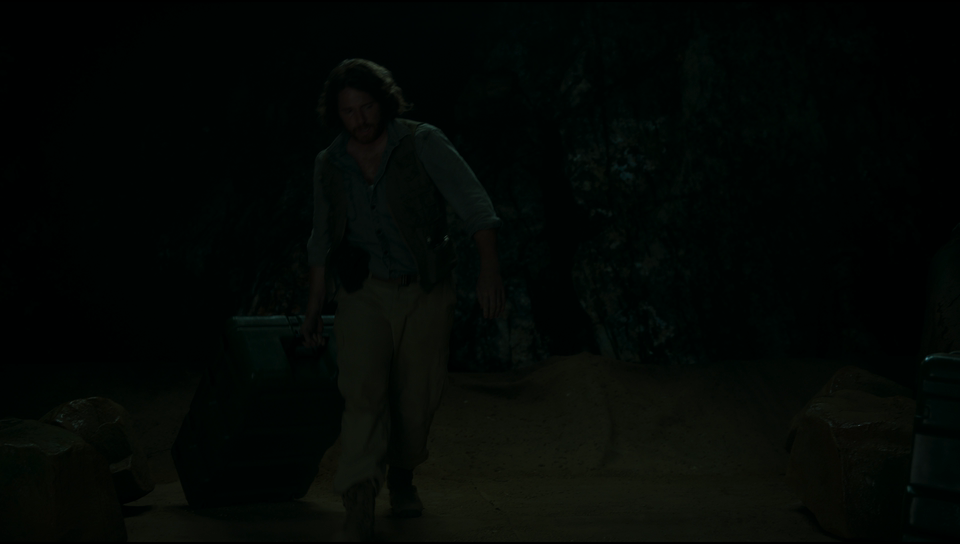}}
        \text{BVI-AOM:CAscStem2S3}\vspace{.1cm}
    \end{minipage}
    
    \caption{Samples of training content from BVI-AOM \cite{nawala2024bvi} and BVI-DVC \cite{ma2021bvi} datasets.}
    \label{fig:bviexample}
\end{figure*}

The RTVSR model, EDSR\_baseline and CVEGAN have been optimised using the same training database as described below. Apart from the provided LDV3 videos, we also collected original sequences from the BVI-DVC database \cite{ma2021bvi} and BVI-AOM database \cite{nawala2024bvi}. BVI-DVC mainly contains PGC (professionally generated content), which has been employed as a training database for MPEG JVET to optimize neural network-based coding tools for VVC. The thumbnails of five representative sequences from these two datasets are shown in Figure \ref{fig:bviexample}.

All original sequences were encoded using SVT-AV1 version 1.8.0, with five quantization parameter (QP) values (31, 39, 47, 55, and 63). Subsequently, both the compressed sequences and their original counterparts were cropped into 48$\times$48 and 144$\times$144 patches (192$\times$192 for $\times$4 track), respectively, and randomly selected for training. Data augmentation techniques such as rotation and flips were applied here to increase content diversity. This resulted in a total of 92,800 pairs of patches. Based on all the generated training material, we trained a single CNN model for compressed video content with various QPs.

\subsubsection{Training Configuration}

The training of the proposed model consists of two stages. In the first stage, a combined perceptual loss function, as described in \cite{ma2020cvegan}, is employed to optimise the model,

\begin{equation}
    \mathcal{L}_{p} = 0.3 \mathcal{L}_\mathit{L1} + 0.2 \mathcal{L}_\mathit{SSIM} + 0.1 \mathcal{L}_\mathit{L2} + 0.4 \mathcal{L}_\mathit{MS-SSIM}
\end{equation}

The training configurations of the CVEGAN and EDSR\_base model can be found in their original paper \cite{ma2020cvegan, lim2017enhanced}, and they are trained on the same datasets.

In the second stage, a similar knowledge distillation strategy as in  \cite{jiang2023compressing} is utilised, where the pre-trained model in the first stage is considered as the student model, while a pre-trained CVEGAN is used as the teacher model. The total loss $\mathcal{L}_\mathit{total}$ at this stage is given as follows:
\begin{equation}
    \mathcal{L}_\mathit{total} = \alpha\mathcal{L}_\mathit{Lap}(I_\mathit{stu}, I_\mathit{gt}) + \Sigma \mathcal{L}_\mathit{Lap}(I_\mathit{stu}, I_\mathit{tchr}),
    \label{Ltotal}
\end{equation}
where $\mathcal{L}_{Lap}(I_{stu}, I_{gt})$ denotes the original loss between the ground truth $I_{gt}$ and the student model’s prediction $I_{stu}$, and $\alpha$ is a tunable weight, set to 0.1, following \cite{morris2023st}.  $\mathcal{L}_{Lap}(I_{stu}, I_{tchr})$ represents the loss between the student $I_{stu}$ and the teacher’s predictions $I_{tchr}$. Here $\mathcal{L}_{Lap}$ is the Laplacian loss \cite{niklaus2018context}.

\paragraph{\textbf{Results and Discussion}}

Table \ref{tab:results_bvi} summarises the performance of the proposed RTVSR method for the test sequences, compared to the provided anchor results (upsampling by a Lanczos5 filter), EDSR\_baseline, and CVEGAN (teacher model). The average improvement of VMAF against the provided results is up to 4.16, the figure for PSNR-Y is nearly 0.24 dB. Since this challenge mainly focus on VMAF, so our model provide better VMAF performance. By using \cite{VideoAI}, the analysis of our proposed RTVSR model is displayed in Table \ref{tab:ablation_bvi}. As observed in the Table, the processing speed for each frame is 0.8 ms for the $\times 3$ track and 2 ms for the $\times 4$ track. These rapid processing times demonstrate the model's ability to handle high-resolution upscaling tasks with impressive efficiency.

\paragraph{\textbf{Implementation details}}

The coding framework is illustrated in Figure \ref{fig:coding_frame}(b). To generate the training sets, prior to encoding, the original input 1080p video is downsampled by a factor of 3 using a Lanczos5 filter. The SVT-AV1 version 1.8.0 \cite{AV1} serves as the Host Encoder that compresses the low resolution video. At the decoder, when the low resolution video stream is decoded, the proposed RTVSR model is applied to reconstruct the full resolution video content. 
For $\times$4 task, the input is 4K video, downsampled at the encoder and later upsampled at the decoder by a factor of 4.

The employed network was implemented in PyTorch version 1.10 \cite{paszke2019pytorch}. We used the following configuration during training:
Adam optimization \cite{kingma2014adam} with the hyper-parameters: $\beta_{1} = 0.9$ and $\beta_{2} = 0.999$; the batch size of 16; 200 training epochs (100 for both stage one and two); initial learning rate of $10^{-4}$; weight decay of 0.1. The training and evaluation operations were executed on NVIDIA RTX3090.

\subsection{Enhancing Real-Time Compressed Image Super-Resolution with ETDS and Edge-oriented Convolution Block}
\label{sec:megas}

\emph{Jae-Hyeon Lee,
Dong-Hyeop Son,
Ui-Jin Choi} \\
\textit{Megastudy Edu, Republic of Korea}

\vspace{5mm}

The solution is based on ``Real-Time 4K Super-Resolution of Compressed AVIF Images"~\cite{Conde_2024_CVPR}, with the topic ``Enhancing RTSR with ETDS and Edge-oriented Convolutional Blocks". We will refer to this as Enhanced ETDS v1. 

Enhanced ETDS v1 successfully improved super-resolution performance by applying a Feature-Enhanced Module and an Edge-oriented Convolution Block (ECB) to the ETDS \cite{Chao_2023_CVPR}. In this challenge, we introduce Enhanced ETDS v2, aimed at improving the inference speed for real-time video super-resolution.

We improved the architecture of Enhanced ETDS v1 for real-time video super-resolution. To increase the model's inference speed, we reduced the input image resolution by half using a convolutional layer. Additionally, the number of blocks in both the Backbone branch and the Residual branch was reduced from 5 to 3, while the number of channels was increased from 24 to 36. Given the shallow architecture, we employed multi-stage training.

\begin{table}
    \centering
    \begin{tabular}{c c c c c c c c}
         \toprule
         Method & \# Params. (M) & FLOPs (G) & Runtime (ms) &  SR ratio & GPU \\
         \midrule
         Enhanced ETDS v1 & 0.0401  & 2511 &  429 &  x3 & A100 \\
         Enhanced ETDS v2 & 0.1366 & 2134 & 258 & x3 & A100 \\
         Enhanced ETDS v1 & 0.0401  & 2511 &  430 &  x4 & A100 \\
         Enhanced ETDS v2 & 0.1366 & 2134 & 258 & x4 & A100 \\
         \bottomrule
    \end{tabular}
    \vspace{2mm}
    \caption{Ablation study of ETDS. The inference speed measurement results are calculated on videoclip of 30 frames.}
    \label{tab:results_megas}
\end{table}

Our method uses a dataset consisting of 1000 samples drawn from DIV2K \cite{Agustsson_2017_CVPR_Workshops}, Flickr2K \cite{Agustsson_2017_CVPR_Workshops}, and LSDIR \cite{Li_2023_CVPR}. The dataset was degraded using random AVIF compression factors between 10 and 90, as well as bicubic interpolation with scaling factors of 3x and 4x. During training, the images were normalized to a range of [0, 1], and image augmentation techniques such as random cropping, flipping, and rotation were applied.

\begin{figure*}[t]
    \centering
    \includegraphics[width=1\textwidth]{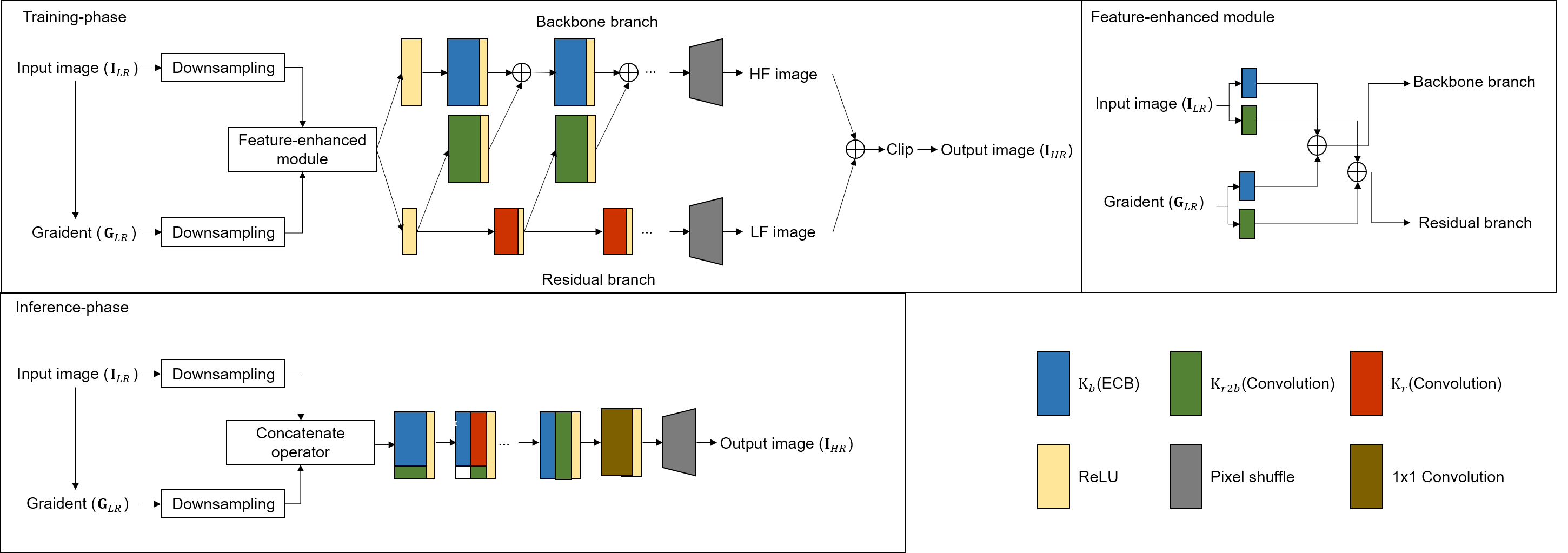}
    \caption{Enhanced ETDS v2 proposed by Team Megastudy.}
    \label{fig:megas}
\end{figure*}

\paragraph{Implementation details}

In the multi-stage training, during the first stage, the model was trained for 800 epochs with an initial learning rate of 1e-5, which gradually decreased to 1e-8. We employed Adam optimizer with parameters $\beta_1 = 0.9$ and $\beta_2 = 0.999$. In the second stage, the model was trained for 1000 epochs with an initial learning rate of 1e-6, decreasing to 1e-9. The model is trained for 48 hours.

For 4x super-resolution, the low-resolution (LR) patch size was set to 64, and the high-resolution (HR) patch size was set to 512, while for 3x super-resolution, the LR patch size was 64 and the HR patch size was 384. The mini-batch size was set to 64. Charbonnier loss function and Adam optimizer was used, with a cosine scheduler for learning rate adjustment.

\subsection{A Simple Feature Modulation Approach for Efficient Video Super-Resolution}
\label{sec:vpeg}

\emph{Mingjun Zheng, Long Sun, Jinshan Pan, Jiangxin Dong, Jinhui Tang} \\
\textit{Nanjing University of Science and technology (VPEG-VSR)}

\vspace{5mm}

\begin{figure*}[t]
    \centering
    \includegraphics[width=1\textwidth]{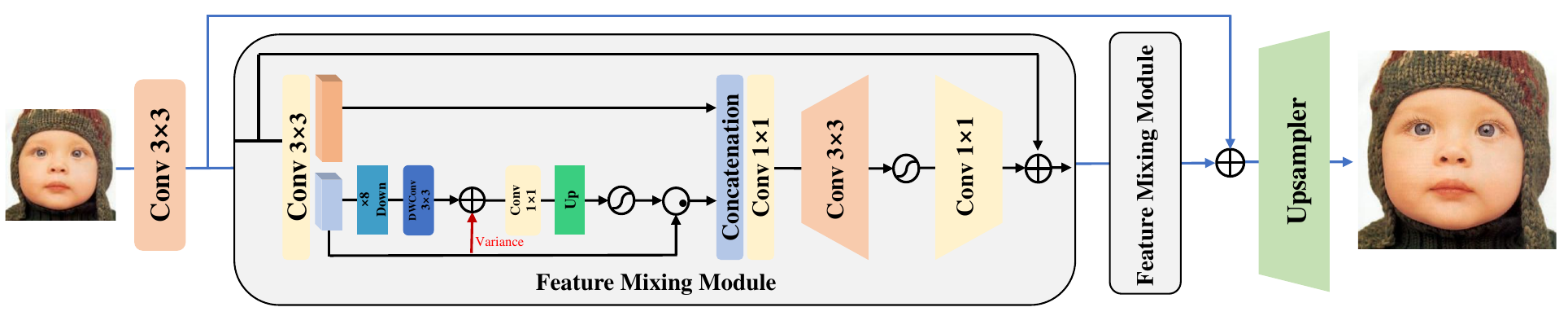}
    \caption{An overview of the proposed model by VPEG-VSR.}
    \label{fig:vpeg}
\end{figure*}

We present a simple feature modulation method for efficient video super-resolution that is a straightforward modification of the SAFMN++~\cite{sun2023safmn}. As shown in figure~\ref{fig:vpeg}, the network consists of a variance-conditional feature modulation block and a CCM layer~\cite{sun2023safmn}. We train the super-resolution model with $\times3$ and $\times4$ enlarging factors on the first 500 clips of the Inter4K~\cite{Inter4K} dataset.

\begin{table}
    \centering
    \setlength{\tabcolsep}{10pt}
    \caption{Summary of VPEG-VSR results on both tracks.}
    \begin{tabular}{c c c c c c c}
        \toprule
         Method &Params & FLOPs & Runtime & PSNR & SSIM & VMAF  \\
         \midrule
        $\times 4$ Ours & 77.51K  & 39.99G & 8.56ms & 31.53  & 0.9111& 36.64\\
       
        $\times 3$ Ours & 70.70K & 16.20G & 3.84ms & 28.84  & 0.8634 & 34.44\\
        \bottomrule
    \end{tabular}

    \label{tab:results}
\end{table}

We introduce a simple feature modulation method for efficient video super-resolution that is a straightforward modification of the SAFMN++~\cite{sun2023safmn}.
Unlike the previous SAFM++, as shown in Fig~\ref{fig:vpeg} the improved module adds global variance as a condition for better feature modulation.
Within this module, a 3$\times$3 convolution is first utilized to extract local features and a single scale feature modulation is then applied to a portion of the extracted features for non-local feature interaction.
After this process, these two sets of features are aggregated by channel concatenation and fed into a 1$\times$1 convolution for feature fusion.
Subsequently, the fused features are fed to the CCM for further processing.

\paragraph{Implementation details}

The proposed video SR model is trained by minimizing a combination of L1 loss and the FFT-based L1 loss~\cite{MIMO} with Adam optimizer for a total of 200,000 iterations. We set the initial learning rate to $1\times10^{-3}$ and the minimum one to $1\times10^{-7}$, which is updated by the Cosine Annealing scheme\cite{cosine}.

We train the super-resolution model with $\times3$ and $\times4$ enlarging factors on the first 500 clips of the Inter4K~\cite{Inter4K} dataset. The cropped LR frame size is 120$\times$120 and the mini-batch size is set to 32.  The training process takes about 24 hours.

\subsection{A Simple Learnable Guided Filter Feature Modulation Approach for Efficient Video Super-Resolution}
\label{sec:nanjing}

\emph{Zhongbao Yang, Long Sun, Jinshan Pan, Jiangxin Dong, Jinhui Tang} \\
\textit{Nanjing University of Science and technology}

\vspace{5mm}

\begin{figure*}[t]
    \centering
    \includegraphics[width=1\textwidth]{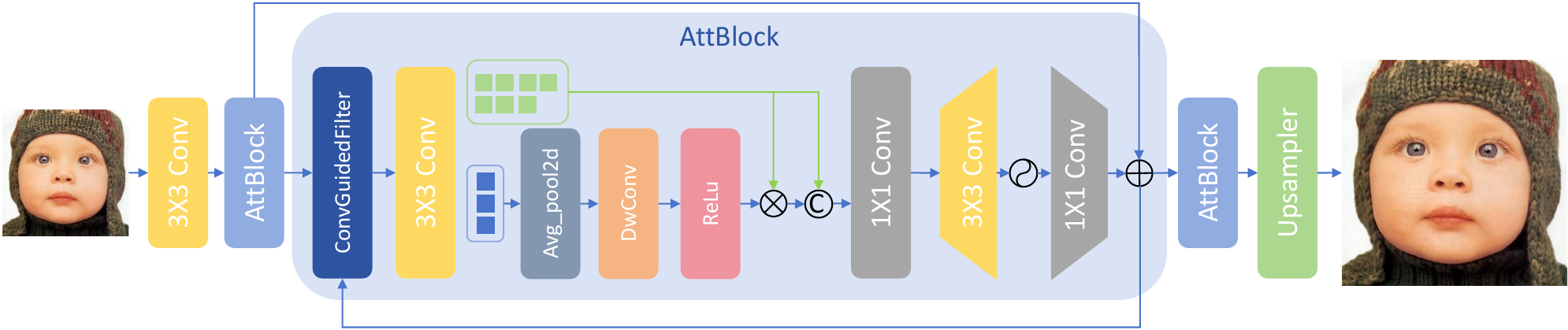}
    \caption{An overview of the proposed modification of SAFMN++~\cite{sun2023safmn}.}
    \label{fig:framework_nanjing}
\end{figure*}

We present a simple learnable guided filter feature modulation method for efficient video super-resolution that is a straightforward modification of the SAFMN++~\cite{sun2023safmn}. As shown in figure~\ref{fig:framework_nanjing}, the network consists of a learneable guided filter~\cite{DBLP:conf/cvpr/Wu0ZH18}, variance-conditional feature modulation block and a CCM layer~\cite{sun2023safmn}. We train the super-resolution model with $\times4$ enlarging factors on the first 500 clips of the Inter4K~\cite{Inter4K} dataset.

Unlike the previous SAFM++, as shown in Fig~\ref{fig:framework_nanjing} the improved module adds learnable guided filter~\cite{DBLP:conf/cvpr/Wu0ZH18} as a condition for better feature modulation. After this process, these two sets of features are aggregated by channel concatenation and fed into a 1$\times$1 convolution for feature fusion. Subsequently, the fused features are fed to the CCM for further processing.

\paragraph{Implementation details}

The model has 40.45K parameters; it requires 20.44G FLOPs and 8.2ms to process one frame from 540p to 4K.

The proposed model is trained by minimizing a combination of L1 loss and the FFT-based L1 loss~\cite{MIMO} with Adam optimizer for a total of 200,000 iterations. 
We set the initial learning rate to $5\times10^{-4}$ and the minimum one to $1\times10^{-7}$, which is updated by the Cosine Annealing scheme~\cite{cosine}.


\section*{Acknowledgements}
This work was partially supported by the Humboldt Foundation. We thank the AIM 2024 sponsors and the challenge sponsors: Meta Reality Labs, Meta, KuaiShou, Huawei, Sony Interactive Entertainment, Netflix Inc., and the University of W\"urzburg (Computer Vision Lab).



\bibliographystyle{splncs04}
\bibliography{main}

\end{document}